\documentclass[aps,prb,twocolumn,superscriptaddress,citeautoscript]{revtex4}

\usepackage{amsmath,amssymb}
\usepackage{bm}
\usepackage{graphicx}
\usepackage{epstopdf}
\usepackage{latexsym}
\usepackage{subfigure}
\usepackage{color}
\usepackage{dsfont} 
\usepackage{wasysym} 
\usepackage{hyperref} 

\newcommand{\be}{\begin{equation}}
\newcommand{\ee}{\end{equation}}
\newcommand{\bea}{\begin{eqnarray}}
\newcommand{\eea}{\end{eqnarray}}

\newcommand{\yto}{{Yb$_2$Ti$_2$O$_7$}}

\begin{document}

\title{Spin Liquid Regimes at Nonzero Temperature in Quantum Spin Ice
}

\author{Lucile Savary}
\affiliation{Department of Physics, University of California, Santa Barbara, California 93106-9530, U.S.A.}
\author{Leon Balents}
\affiliation{Kavli Institute for Theoretical Physics, University of
  California, Santa Barbara, California, 93106-4030, U.S.A.}

\date{\today}
\begin{abstract}

Quantum spin liquids are highly entangled ground states of quantum systems with an emergent gauge structure, fractionalized spinon excitations, and other unusual properties.  While these features clearly distinguish quantum spin liquids from conventional, mean-field-like states at zero temperature ($T$), their status at $T>0$ is less clear.  Strictly speaking, it is known that most quantum spin liquids lose their identity at non-zero temperature, being in that case adiabatically transformable into a trivial paramagnet.  This is the case for the $U(1)$ quantum spin liquid states recently proposed to occur in the quantum spin ice pyrochlores.  Here we propose, however, that in practical terms, the latter quantum spin liquids can be regarded as phases distinct from the high temperature paramagnet.  Through a combination of gauge mean field theory calculations and physical reasoning, we argue that these systems sustain both quantum spin liquid and {\em thermal} spin liquid phases, dominated by quantum fluctuations and entropy, respectively.  These phases are separated by a first order ``thermal confinement'' transition such that, for temperatures below the transition, spinons and emergent photons are coherently propagating excitations, and above it the dynamics is classical.  Even for parameters for which the ground state is magnetically ordered and not a quantum spin liquid, this strong first order transition occurs, pre-empting conventional Landau-type criticality.  We argue that this picture explains the anomalously low temperature phase transition observed in the quantum spin ice material Yb$_2$Ti$_2$O$_7$.  
\end{abstract}

\maketitle

\section{Introduction}
\label{sec:introduction}

Since the discovery of the quantum Hall effect, it has been recognized that phases of matter at zero temperature can be distinguished by means other than symmetry, e.g. in that case by a quantized Hall conductance, related to a topological invariant.\cite{prange2012quantum}\  In recent years, the quantum Hall state has been placed into a much broader class of phases with topological order of various types, or more generally quantum order or long range entanglement.\cite{wen2007quantum}  Theoretically, many such states may be realized as Quantum Spin Liquids (QSLs): ground states of frustrated quantum magnets with long range entanglement (and usually, though not necessarily, the absence of symmetry breaking).\cite{balents2010}  The key feature of such QSL states is that they support non-trivial excitations which cannot be created individually by any local operator, and with  mutual statistics (and often quantum numbers) different from that of the bare electron and composites made from it.  In some cases, a quantitative measure of the long range nature of the entanglement can be devised through a study of the entanglement entropy, a non-local quantity defined in terms of the ground state.\cite{kitaev2006topological,levin2006detecting}  The entanglement entropy is very difficult to measure experimentally, however.

While {\em zero temperature} ($T$) phases can be characterized and distinguished by entanglement properties and by their excitations, these criteria fail at nonzero temperature, the former for obvious reasons.  The excitation criteria is also invalid at $T>0$,  where the description of the system involves the thermal superposition of all eigenstates of the Hamiltonian. In the thermodynamic limit, even if the {\em  density} of elementary excitations (provided they can even be defined) is small and finite, their number is infinite, and the description of an ``excitation above the thermal ground state'' makes little sense.   For a system with a QSL ground state, which lacks a symmetry-based characterization, one is tempted to conclude that at any non-zero temperature, there is no qualitative distinction of the physics from that of a trivial paramagnet, and consequently no phase transition on heating from absolute zero to high temperature.  

The true answer is more complex.  Some understanding may be gleaned from work on lattice gauge theory, a natural framework for QSL states.  It is known that, in some but not all cases, the low and high temperature phases {\em  can} be qualitatively distinguished, so that a phase transition at non-zero $T$ is {\em inevitable}.  For example it is known that the Coulomb $U(1)$ spin liquid in three spatial dimensions and the $\mathbb{Z}_2$ spin liquid in two spatial dimensions need not go through a transition as the temperature is increased, while the $\mathbb{Z}_2$ phase in three spatial dimensions at infinitesimal temperature is not qualitatively the same as the paramagnetic phase.\cite{svetitsky1986symmetry}   This difference can be understood from a picture of the topological defects of the QSL states, whose proliferation coincides with and is necessary for the complete destruction of the spin liquid state.  In the 2d $\mathbb{Z}_2$ and 3d $U(1)$ QSLs, the topological defects are {\em  point-like} objects, $\mathbb{Z}_2$ vortices or ``visons'' in the former case and magnetic/electric monopoles in the latter case.  The pointlike defects are always created with non-zero density at $T>0$, making these states smoothly connected to a paramagnet.  By contrast, in the 3d $\mathbb{Z}_2$ QSL, the defects are $\mathbb{Z}_2$ vortex {\em lines}. In the form of small loops, such lines do not disrupt the QSL state. The latter is therefore only destroyed when infinitely long loops proliferate, but, due to their infinite total energy (non-zero energy per unit length), they do so only above a non-zero critical temperature.  

The above considerations give only some indication of the presence or absence of a phase transition but do not prejudge of the existence of conventional first order transitions in cases where one is not {\em  required}.  This is a non-universal question, a type often frowned upon by theorists.  However, this non-universality has the virtue that the answer may shed light on the microscopic physics of the system.  In this paper, we address it in the specific case of the general model for nearest-neighbor quantum pyrochlore antiferromagnets studied in Refs.\onlinecite{theorypaper,jointpaper}, which contains a $U(1)$ QSL phase at zero temperature, and which can be considered a concrete model for ``quantum spin ice''.  The model and the question of a thermally driven phase transition is particularly relevant to the case of \yto, which appears in some experiments to exhibit a QSL ground state and also clearly shows a sharp phase transition in the best quality samples.\cite{blote1969,hayre2012,chang2012,ross2009,gardner2004,hodges2002,thompson2011}    While on the general grounds mentioned above there is no requirement for a $T>0$ phase transition, and indeed there is none in the simple $U(1)$ lattice gauge theories, we argue that the quantum spin ice model model does indeed exhibit a first order phase transition.  The difference is that quantum spin ice is described by a $U(1)$ gauge theory {\em  strongly coupled} to ``spinons'', fractional spin excitations which appear in the theory as matter fields carrying the $U(1)$ gauge charge.  Thus in quantum spin ice the matter matters. 

The first order transition from the QSL to the high temperature phase occurs without any change of symmetry, and is thus an analog of the liquid-gas transition.  Indeed, pushing this analogy, we argue that the transition may be regarded as a catastrophic collapse of the QSL state occuring at $T<T_c$, which is supported by quantum coherence, to a {\em  thermal} spin liquid state for $T>T_c$, supported instead by a large residual entropy.  In fact, the tendency of the thermal spin liquid to supplant the quantum coherent phases is so strong that the collapse transition persists even in regions where the ground state is {\em not} a QSL, but an ordered ferromagnet or antiferromagnet (which appear as Higgs phases in the theory).  We will return to the implications of this finding for the putative Higgs transition observed in \yto\ at the end of this paper.

\begin{figure}[htbp]
\begin{center}
\includegraphics[width=3.3in]{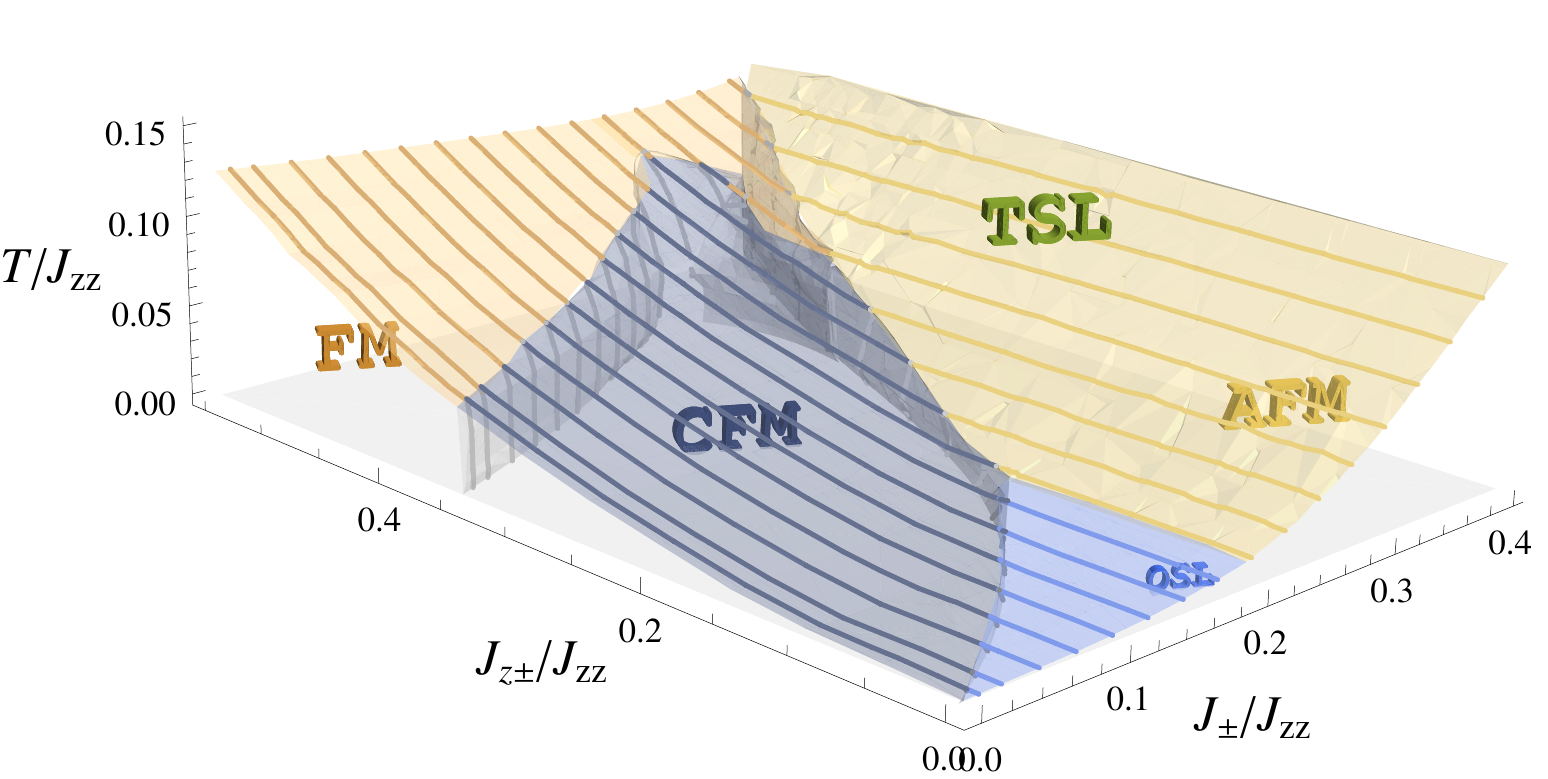}
\caption{(Color online) Finite temperature gauge mean field phase diagram obtained for $J_{\pm\pm}=0$ and $J_{zz}>0$.  ``QSL'', ``CFM'', ``FM'', ``AFM'' and ``TSL'' denote the $U(1)$ Quantum Spin Liquid, Coulomb Ferromagnet, standard ferromagnet, standard antiferromagnet, and Thermal Spin Liquid, respectively. The lines represent fixed-$J_\pm/J_{zz}$ cuts and are highlighted here to serve as guides to the eye. Details of how this figure was obtained are given in Appendix~\ref{sec:figures-calc}.}
\label{fig:MF-3dphasediag}
\end{center}
\end{figure}

Our results are summarized by the three-dimensional phase diagram shown in Fig.~\ref{fig:MF-3dphasediag}, which includes two exotic phases, namely a $U(1)$ Coulomb quantum spin liquid (QSL) and a $U(1)$ ``Coulombic ferromagnet'' (CFM), whose properties are now well known at zero temperature.\cite{theorypaper,lee2012}  Our calculations extend the zero temperature diagram in the $J_{\pm}/J_{zz}-J_{z\pm}/J_{zz}$ plane to include the temperature axis, $T/J_{zz}$, and as advertised above this diagram contains a large-entropy thermal spin liquid phase (TSL), in addition to the zero temperature like phases.   The boundaries in Fig.~\ref{fig:MF-3dphasediag} are calculated using the extension of gauge Mean Field Theory (gMFT), described below.  While one may be concerned about possible artifacts due to this approximation, we provide physical arguments that phase boundaries we obtain are qualitatively correct.  
Notably, the transition to the TSL is first order and occurs at a temperature strikingly lower than the natural energy scales such as the exchange couplings themselves and the Curie-Weiss temperature.  Indeed, in the perturbative regime with $J_\pm \ll J_{zz}$ and $J_{z\pm}=0$ studied by Hermele {\it et al},\cite{hermele2004} analytic arguments imply it occurs at $k_B T_c \sim J_{\pm}^3/J_{zz}^2 \ll J_{zz}$.  The gMFT approximation actually overestimates $T_c$ in this limit, giving $k_B T_{c}^{\rm gMFT} \sim J_{\pm}^2/J_{zz}$, but does qualitatively capture its smallness relative to natural energy scales.

\begin{figure*}[htbp]
\begin{center}
\includegraphics[width=\linewidth]{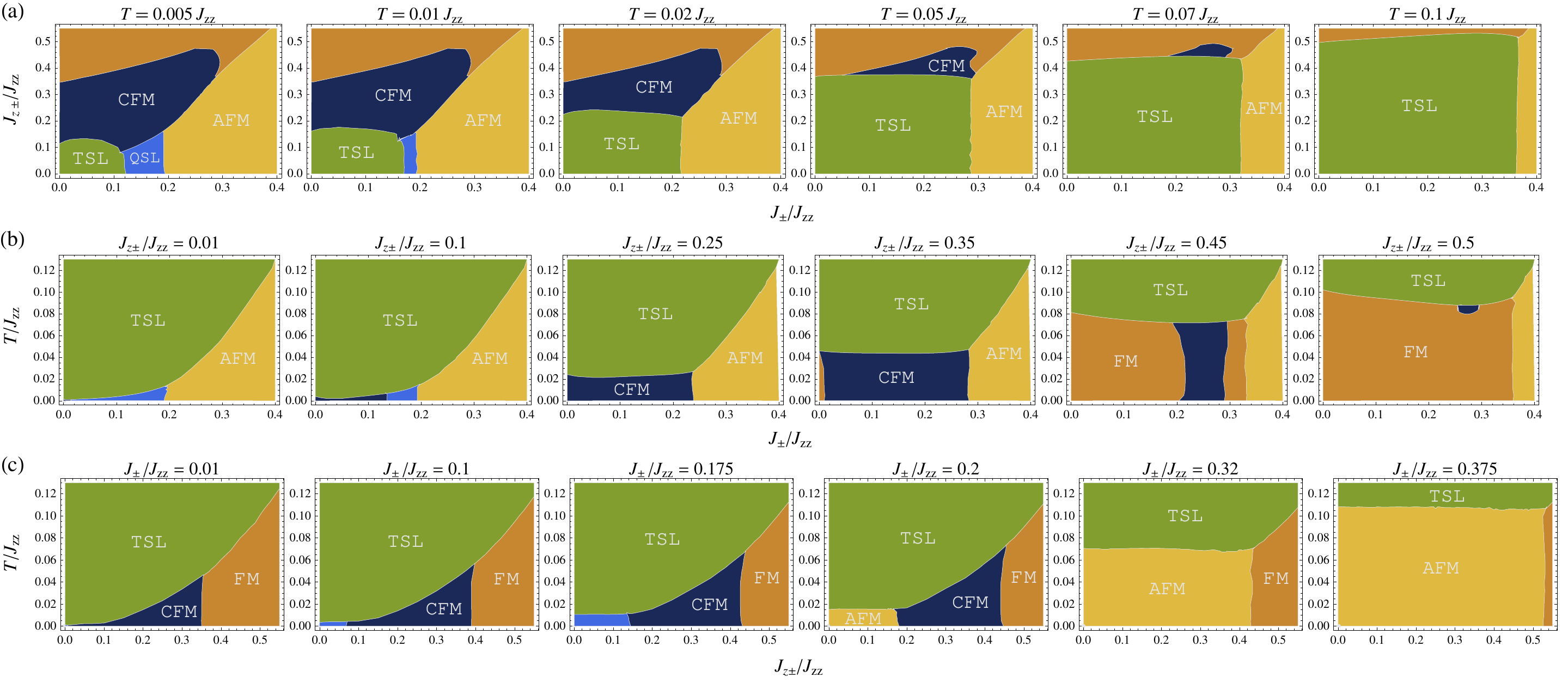}
\caption{(Color online) Cuts through the three-dimensional finite-temperature gauge mean field phase diagram obtained for $J_{\pm\pm}=0$ and $J_{zz}>0$. ``QSL'', ``CFM'', ``FM'', ``AFM'' and ``TSL'' denote the $U(1)$ Quantum Spin Liquid, Coulomb Ferromagnet, standard ferromagnet, standard antiferromagnet, and Thermal Spin Liquid, respectively. Subfigure~(a) (resp. (b), (c)) shows cuts for fixed values of $T/J_{zz}$ (resp. $J_{z\pm}/J_{zz}$, $J_{\pm}/J_{zz}$). Details of how this figure was obtained are given in Appendix~\ref{sec:figures-calc}.}
\label{fig:threecuts}
\end{center}
\end{figure*}

We proceed as follows. We first set up $T>0$ gMFT, mostly extending the analysis introduced in Ref.~\onlinecite{theorypaper}, and present our results after describing the methods used to obtain the three-dimensional phase diagram of Figure~\ref{fig:MF-3dphasediag}. Finally, we discuss our results in the context of the QSL candidate Yb$_2$Ti$_2$O$_7$, whose Hamiltonian is known quantitatively.

\section{Gauge theory}
\label{sec:gauge-theory}

\subsection{Formulation}
\label{sec:gauge-formulations}

In this section, we recapitulate the spin Hamiltonian and its exact slave particle reformulation introduced in Ref.~\onlinecite{theorypaper}.  The Hamiltonian of the system is 
\begin{eqnarray}
  \label{eq:1}
  H & = & \sum_{\langle ij\rangle} \Big[ J_{zz} \mathsf{S}_i^z \mathsf{S}_j^z - J_{\pm}
  (\mathsf{S}_i^+ \mathsf{S}_j^- + \mathsf{S}_i^- \mathsf{S}_j^+) \nonumber \\
  && +\, J_{\pm\pm} \left[\gamma_{ij} \mathsf{S}_i^+ \mathsf{S}_j^+ + \gamma_{ij}^*
    \mathsf{S}_i^-\mathsf{S}_j^-\right] \nonumber \\
  && +\, J_{z\pm}\left[ \mathsf{S}_i^z (\zeta_{ij} \mathsf{S}_j^+ + \zeta^*_{ij} \mathsf{S}_j^-) +
    {i\leftrightarrow j}\right]\Big],
\end{eqnarray}
where the sans serif characters $\mathsf{S}_i^\mu$ denote components of
the spins in the {\em local} pyrochlore bases, where $\gamma$ is a
$4\times4$ matrix with only non-zero off-diagonal entries, which are
complex unimodular numbers, and $\zeta_{ij}=-\gamma_{ij}^*$
\cite{jointpaper,theorypaper} whose explicit expression as well as those
of the local bases used in Eq.~\eqref{eq:1} are given in
Appendix~\ref{sec:notations}.  Details of the gMFT formalism were given
in Ref.~\onlinecite{theorypaper} and we give here the main definitions
and results. The gauge ``charge'' on each diamond site is
\begin{equation}
\label{eq:2}
Q_\mathbf{r}=\eta_\mathbf{r}\sum_\mu\mathsf{S}^z_{\mathbf{r},\mathbf{r}+\eta_\mathbf{r}\mathbf{e}_\mu},
\end{equation}
where $\eta_\mathbf{r}=1$ (resp. $-1$) for a I (resp. II) diamond sublattice site, and the $\mathbf{e}_\mu$ are the four nearest-neighbors of an $\eta_\mathbf{r}$ (I) diamond sublattice site. The Hilbert space is enlarged, and the spins are rewritten
\begin{equation}
  \label{eq:3}
  \mathsf{S}_{\mathbf{r},\mathbf{r}+\mathbf{e}_\mu}^+ =
\Phi_{\mathbf{r}}^\dagger\,
\mathsf{s}_{\mathbf{r},\mathbf{r}+\mathbf{e}_\mu}^+
\Phi_{\mathbf{r}+\mathbf{e}_\mu}, \qquad
\mathsf{S}_{\mathbf{r},\mathbf{r}+\mathbf{e}_\mu}^z =
\mathsf{s}_{\mathbf{r},\mathbf{r}+\mathbf{e}_\mu}^z,
\end{equation}
which will allow implementation of mean field theory, while preserving the possibility for the description of exotic phases. Here $\Phi_\mathbf{r}=e^{-i\varphi_\mathbf{r}}$ is a bosonic spinon field, with $[\varphi_\mathbf{r},Q_\mathbf{r}]=i$. For $J_{\pm\pm}=0$, within this exact reformulation the Hamiltonian becomes a nearest- and next-nearest-neighbor hopping Hamiltonian for the spinons, in a background of fluctuating gauge fields (see Eq.~(4) of Ref.~\onlinecite{theorypaper}). This Hamiltonian is invariant under the $U(1)$ gauge transformation
\begin{equation}
\label{eq:gauge-sym}
\begin{cases}
\Phi_\mathbf{r}\rightarrow\Phi_\mathbf{r}\,e^{-i\chi_\mathbf{r}}\\
{\sf s}^\pm_{\mathbf{rr}'}\rightarrow {\sf s}_{\mathbf{rr}'}^\pm e^{\pm i(\chi_{\mathbf{r}'}-\chi_{\mathbf{r}\vphantom{'}})}
\end{cases},
\end{equation}
for any arbitrary real function $\mathbf{r}\mapsto\chi_\mathbf{r}$. 

\subsection{Gauge mean field theory at $T>0$}
\label{sec:gauge-mean-field}

Ref.~\onlinecite{theorypaper} introduced a ``gauge mean field theory'' (gMFT) at $T=0$ to decouple the matter fields from the gauge fields.  Here we extend this analysis to $T>0$. Following Ref.~\onlinecite{theorypaper} we now make the Ansatz
\begin{equation}
\langle\mathsf{s}_\mu^z\rangle=\mathsf{s}\sin\theta\,\varepsilon_\mu,\qquad\langle\mathsf{s}^-_\mu\rangle=\mathsf{s}\cos\theta,
\label{eq:ansatz}
\end{equation}
where the expectation value of an operator $U$ is $\langle
U\rangle=\frac{1}{Z}{\rm Tr}\left[U\, e^{-\beta H}\right]$, with
$\beta=1/(k_B T)$ the inverse temperature for Boltzmann's constant $k_B$
and $Z={\rm Tr}e^{-\beta H}$ the partition function, $\mu=0,..,3$ and
$\varepsilon=(1,1,-1,-1)$. This Ansatz assumes translational invariance
(as seen in experiment), and is compatible with FM polarization along
the (global) $x$ axis
($\langle\mathsf{s}^+_\mu\rangle=\langle\mathsf{s}^-_\mu\rangle$). $0\leq\mathsf{s}=|\langle\vec{\mathsf{s}}\rangle|\leq1/2$
represents the magnitude of the expectation value of the spin, which we
know in two limits, $\mathsf{s}(T=0)=1/2$ (as in
Ref.~\onlinecite{theorypaper}) and $\mathsf{s}(T=+\infty)=0$.  The
variation of the magnitude $\mathsf{s}$ (and in particular the
possibility for it to be zero) is the key new ingredient for $T>0$. 

To avoid spurious solutions, here we do not solve the consistency equations but rather calculate and minimize the variational free energy
\begin{equation}
F_v=F_0+\langle H-H_0\rangle_0,
\end{equation}
where now $F_0$ is defined to be $Z_0=e^{-\beta F_0}$, where $Z_0={\rm Tr}\,e^{-\beta H_0}$ is the partition function of a fiducial system (we know that the extrema of the variational free energy should be the solutions of the mean field consistency equations).\footnote{We thank Matthew Hastings for pointing out this formula to us.} The expectation value of an operator $U$ with respect to the trial Hamiltonian $H_0$ is defined to be as usual $\langle U\rangle_0=\frac{1}{Z_0}{\rm Tr}\left[U\,e^{-\beta H_0}\right]$.  As trial Hamiltonian, we choose a decoupled one made of a sum of a free spin Hamiltonian and that of a simple nearest-neighbor and next-nearest-neighbor hopping Hamiltonian on the diamond lattice, i.e. $H_0=H_\Phi^0+H_\mathsf{s}^0$ with
\begin{widetext}
\begin{eqnarray}
\label{eq:h0phi}
  H_\Phi^0   &=&  \sum_{\mathbf{r} \in {\rm I},{\rm II}} \frac{J}{2} Q_\mathbf{r}^2   - \left\{\sum_{\mathbf{r}\in {\rm I}}  \sum_{\mu,\nu\neq\mu} t'_{\mu\nu}\, \Phi_{\mathbf{r}+\mathbf{e}_\mu}^\dagger
      \Phi_{\mathbf{r}+\mathbf{e}_\nu}^{\vphantom\dagger}
      +
       \sum_{\mathbf{r}\in
        {\rm II}}  \sum_{\mu,\nu \neq \mu} {t'_{\mu\nu}}^*\, \Phi_{\mathbf{r}-\mathbf{e}_\mu}^\dagger \Phi_{\mathbf{r}-\mathbf{e}_\nu}^{\vphantom\dagger} 
    \right\} \\
  && - \left\{ \sum_{\mathbf{r}\in {\rm I}} \sum_{\mu} \left(
    t_\mu\,\Phi_\mathbf{r}^\dagger\, \Phi_{\mathbf{r}+\mathbf{e}_\mu}^{\vphantom\dagger}
    + {\rm h.c.}\right) 
+ 
  \sum_{\mathbf{r}\in {\rm II}} \sum_{\mu} \left(
    t_\mu\,  \Phi_{\mathbf{r}-\mathbf{e}_\mu}^\dagger \Phi_{\mathbf{r}}^{\vphantom\dagger}
    + {\rm h.c.}\right) \right\},
\nonumber
\end{eqnarray}
\end{widetext}
and
\begin{equation}
\label{eq:h0s}
H_\mathsf{s}^0=-\sum_{\mathbf{r}\in{\rm I}}\sum_\mu \vec{\mathsf{h}}_\mu(\mathbf{r})\cdot\vec{\mathsf{s}}_{\mathbf{r},\mathbf{r}+\mathbf{e}_\mu},
\end{equation}
where $J$, $t'_{\mu\nu}$, $t_\mu$, $\vec{\mathsf{h}}_\mu$ are real variational parameters, which we constrain below.

We need to determine the free energy $F_0=-\frac{1}{\beta}\ln Z_0$.  The free spin part $F^0_\mathsf{s}$ is trivial, and $F_\Phi^0$ is obtained as described in Ref.~\onlinecite{theorypaper} and Appendix~\ref{sec:details} (see in particular Eqs.~\eqref{eq:frees} and \eqref{eq:freephi}).  Now, trying $J=J_{zz}$, $t'_{\mu\nu}=t'=J_\pm\mathsf{s}^2\cos^2\theta$, and $t_\mu=\varepsilon_\mu t=\varepsilon_\mu J_{z\pm}\mathsf{s}^2\sin2\theta$ in our variational wavefunction, we get
\begin{eqnarray}
\label{eq:freeen}
&&F_v/N_{u.c.}=\\
&&2\left\{2T\left[\left(\tfrac{1}{2}+\mathsf{s}\right)\ln\left(\tfrac{1}{2}+\mathsf{s}\right)+\left(\tfrac{1}{2}-\mathsf{s}\right)\ln\left(\tfrac{1}{2}-\mathsf{s}\right)\right]-\lambda\right\}\nonumber\\
&&\qquad+\frac{1}{N_{u.c.}}\sum_\mathbf{k}\sum_{i=\pm1}\left\{\omega_{\mathbf{k}}^i-2T\ln\frac{1}{1-e^{-\beta\omega_{\mathbf{k}}^i}}\right\},\nonumber
\end{eqnarray}
where $\omega_{\mathbf{k}}^\pm=\sqrt{2J_{zz}}\sqrt{\lambda-\tilde{L}_\mathbf{k}\pm|\tilde{M}_\mathbf{k}|}$, $\tilde{L}_\mathbf{k}=\frac{J_1}{2}\sum_{\mu,\nu\neq\mu}\cos\left(\mathbf{k}\cdot(\mathbf{e}_\mu-\mathbf{e}_\nu)\right)$, $\tilde{M}_\mathbf{k}=J_2\sum_{\mu}\varepsilon_\mu e^{i\mathbf{k}\cdot\mathbf{e}_\mu}$ with $J_1=2J_\pm\mathsf{s}^2\cos^2\theta$, $J_2=2J_{z\pm}\mathsf{s}^2\sin2\theta$, $\varepsilon=(1,1,-1,-1)$. We will also be using $
L_\mathbf{k}=\frac{1}{2}\sum_{\mu,\nu\neq\mu}\cos\left(\mathbf{k}\cdot(\mathbf{e}_\mu-\mathbf{e}_\nu)\right)$, $M_\mathbf{k}=\sum_{\mu}\varepsilon_\mu e^{i\mathbf{k}\cdot\mathbf{e}_\mu}$. Like in Ref.~\onlinecite{theorypaper}, $\lambda$ is a Lagrange parameter present to enforce the constraint on the spinons (rotor operators) $\Phi_\mathbf{r}^\dagger\Phi^{\vphantom{\dagger}}_\mathbf{r}=1$, in the form $\langle\Phi_\mathbf{r}^\dagger\Phi^{\vphantom{\dagger}}_\mathbf{r}\rangle=1$, i.e.
\begin{equation}
\label{eq:I3}
1=I_3=\frac{1}{2N_{u.c.}}\sqrt{\frac{J_{zz}}{2}}\sum_\mathbf{k}\left[\frac{\mathcal{F}_\mathbf{k}^+}{\sqrt{\lambda-\ell_\mathbf{k}^+}}+\frac{\mathcal{F}_\mathbf{k}^-}{\sqrt{\lambda-\ell_\mathbf{k}^-}}\right],
\end{equation}
where
\begin{equation}
\label{eq:Fcaldef}
\mathcal{F}^\pm_\mathbf{k}=\coth\left[\beta\sqrt{\frac{J_{zz}}{2}}\sqrt{\lambda-\ell_\mathbf{k}^\pm}\right],\qquad\ell^\pm_\mathbf{k}=\tilde{L}_\mathbf{k}\mp|\tilde{M}_\mathbf{k}|.
\end{equation}

In the condensed phases, we find $\lambda=\lambda_{\rm min}+\hat{\delta}\frac{T}{N_{u.c.}}$, where $\lambda_{\rm min}=\max_\mathbf{k}\ell_\mathbf{k}^-$ and $\hat{\delta}=O(1)$, with $\hat{\delta}$ positive and independent of $T$ (the difference in the exponent compared with the zero temperature case explored in Ref.~\onlinecite{theorypaper} is addressed in Appendix~\ref{sec:I3}).

Taking the $T\rightarrow0$ limit of $F_v$ and comparing with the ground state energy found at zero temperature is rather subtle, and carefully described in Appendix~\ref{sec:zerotemp}.  We find $\lim_{T\rightarrow0}F_v=-2\lambda(T=0)+\sum_{i=\pm1}\omega^i(T=0)$, where $\lambda(T=0)$ is determined by the $I_3(T=0)=1$ equation at zero temperature, and which is a variational form of the zero-temperature ground state energy (see Appendix~\ref{sec:zerotemp}).

\section{Results}
\label{sec:phase-diagram}

\subsection{Phase diagram}
\label{sec:phase-diagram-1}

We find the phase diagram presented in Figure~\ref{fig:MF-3dphasediag}. It contains the ``continuation'' of the four phases present at zero temperature: the conventional ferromagnet and antiferromagnet, the deconfined $U(1)$ quantum spin liquid and Coulomb ferromagnet (CFM). This diagram is also enriched by the additional thermal spin liquid (TSL) phase mentioned in the introduction, which exists at temperatures $T_c \leq T\ll J_{zz}$.

Within gMFT, we find that the transition to the TSL is first order and, at small $J_\pm/J_{zz}$ and $J_{z\pm}/J_{zz}$, occurs when $k_BT_c\sim J_i^2/J_{zz}^2$.  More precisely
\begin{equation}
T_cJ_{zz}=\frac{3\cos^4\theta}{16\ln2}J_\pm^2+\frac{\sin^22\theta}{4\ln2}J_{z\pm}^2,
\end{equation}
where $\theta$ needs to have been well chosen to minimize the $T=0$ energy (see Appendix~\ref{sec:smallparams}).
Using the perturbative limit of the theory\cite{jointpaper}, the transition temperature would be expected to scale as $k_B T_c^{\rm pert}\sim J_\pm^3/J_{zz}^2$.  This means that mean field theory {\em  overestimates} the magnitude of transition temperature.   

The zero-temperature properties of the $U(1)$ spin liquid and CFM were described at length in Refs.~\onlinecite{hermele2004,jointpaper,theorypaper,lee2012}.  The elementary excitations in these phases are deconfined fractional particles -- the spinons and monopoles -- as well as a gapless photon, which arises thanks to fluctuations of the electric field and vector potential.\footnote{Throughout the paper, we use the usual vocabulary of quantum field theory, which is also that used in Ref.~\onlinecite{hermele2004}. This means in particular that our ``electric'' fields are the local $z$-components of the spins (called ``magnetic'' fields in the classical spin ice literature), and that the (classical analogues of the) spinons denote what are referred to as ``{\em  magnetic} monopoles'' in the classical spin ice literature.} Hallmarks of those excitations can in principle be seen in inelastic neutron scattering. The former two appear as a diffuse signal, and the photon as a sharp, linearly dispersing mode whose amplitude vanishes on approaching $\mathbf{k}=\mathbf{0}$. Within the gMFT approach, despite the fluctuations of the gauge and ``electric'' fields, these phases are nevertheless described by a nonzero $\langle\mathsf{s}^-\rangle$, thanks to $\langle\Phi\rangle=0$.  In the mean field sense these phases ``survive'' at low but non-zero temperature, i.e. we retain $\langle\Phi\rangle=0$ and $\langle\mathsf{s}^-\rangle\neq0$, with the latter expectation value reduced in magnitude by thermal fluctuations (in fact the reduction is very small for all temperatures below $T_c$).   This is consistent with the notion that the topologically non-trivial spinon and monopole excitations -- those generating long-range electric and/or magnetic fields -- are dilute at low temperature, due to their non-zero energy cost (gap).  A ``black body'' spectrum of thermally excited artificial photons will also be produced by thermal fluctuations, but as the photons are themselves weakly interacting (they interact only via anharmonic terms whose effects are small at low energy), the thermally excited photons do not induce significant scattering.    Physically, the neutron structure factor should also remain qualitatively similar to its form at zero temperature, modified mainly by small thermal rounding.  

Let us turn to the TSL phase.  Here the gMFT solution is qualitatively changed, and with $\langle{\mathsf{s}}^-\rangle=0$ and likewise $\langle \vec{\mathsf{h}}_\mu\rangle=0$.  The former implies that, at the mean field level, the spinons cannot hop (recall that at least one factor of $\langle\mathsf{s}^\pm\rangle$ enters every spinon hopping amplitude), and the latter implies that the spins $\vec{\mathsf{s}}$ are freely fluctuating thermally.  Consequently we can view the TSL state as one in which spinons are {\em non-propagating}, and where there is a large true entropy.  Physically, this is the best the mean field theory can do to emulate the situation in classical spin ice, in which the spins are completely free apart from the two-in/two-out constraint.  This constraint is itself relaxed slowly as the temperature is raised from well below $J_{zz}$ to well above it.  Hence the TSL state is adiabatically connected to the paramagnetic phase, which is described by the same order parameter values within gMFT.  Indeed, if the $T_c$ becomes sufficiently large, i.e. comparable to $J_{zz}$, as it will deep in the FM or AFM regimes, then the paramagnetic state becomes trivial, and features like pinch points need not arise.  

\subsection{Validity of the gMFT treatment}
\label{sec:valid-gmft-treatm}

Within the gMFT solution, the TSL appears via a strong first order transition from the exotic states at lower temperature.  One may be suspicious of this conclusion, since a mean field treatment of related lattice gauge theories sometimes gives spurious first order transitions.  For example, numerical studies of the simplest pure compact $U(1)$ gauge theory without matter fields show that it undergoes a smooth evolution from $T=0^+$ to high temperature, without any phase transition.  Nevertheless, a mean field treatment predicts a first order transition in that case as well.\cite{savary2013} It is therefore natural to ask: What is wrong with the mean field treatment in this case?  Does the same problem affect our calculations for the pyrochlore problem?

In the pure compact $U(1)$ gauge theory, the Hilbert space consists of the space of electric field configurations on the lattice which {\em strictly} satisfy the charge neutrality constraint ${\rm div}\, {\bf E}=0$ on all sites.  The Hamiltonian in this case has the usual form,
\begin{equation}
  \label{eq:5}
  H_{U(1)} = U\sum_{\langle ij\rangle} \frac{E_{ij}^2}{2} - K \sum_p \cos (\nabla \times A)_p,
\end{equation}
where the second sum is over spatial plaquettes $p$, and gives an explicit microscopic stiffness $K$ penalizing configurations with non-zero magnetic flux $B_p =  (\nabla \times A)_p$. As usual, $A_{ij}$ and $E_{ij}$ are canonically conjugate variables.   The mean field treatment consists of decoupling this stiffness term by defining a self-consistent value of the bond ``order parameter'' $\langle e^{iA_{ij}}\rangle$, calculated using a Hamiltonian for decoupled bonds.  At the mean field level, there are two phases: one where this expectation value $\langle e^{iA_{ij}}\rangle$ is non-zero, defining a putative ``Coulomb phase'', and another where the expectation value vanishes, signaling a ``confined'' phase.  This transition occurs abruptly, i.e. is first order, because the mean field treatment essentially neglects correlations amongst the fluctuations of the electric fields.  In reality, if one considers the situation with large $K$ and low temperature, the electric field fluctuations are highly correlated.  They arise from magnetic monopole excitations, i.e. textures in $B_p$, whose motion leads to long-range electric field fluctuations through Faraday's law.  The complicated spatial structure of magnetic monopoles is completely missed in the mean field treatment.  Instead, in mean field theory, the electric field fluctuations occur locally (and do not even obey the diverge-free constraint since the latter is only satisfied on average in mean field theory).  In fact, because monopoles have a finite energy, at any $T>0$ they appear in a non-zero (albeit exponentially small) concentration, and their motion immediately leads to the destruction of the Coulomb phase, in a strict sense.  Specifically, they screen the interactions between inserted test magnetic charges, and lead to exponential decay of all correlations.  Nevertheless, at very low temperature, because the monopoles are very dilute, the short distance stiffness given by $K$ is largely unaffected.  With increasing temperature, the density of monopoles increases, gradually reducing the effective stiffness on the scale of the correlation length (distance between monopoles).  We can conclude that mean field theory fails in this case because it misses the true mechanism of destruction of the Coulomb phase at $T>0$, which is magnetic monopole proliferation.

Now let us turn to the real pyrochlore problem.  Here the gauge theory is distinguished from the pure $U(1)$ problem described above by the presence of matter (spinon) fields $\Phi_{\bf r}$ (and by the constraint on the magnitude of the electric fields, but this is of secondary importance).  Most importantly, there is in fact no microscopic stiffness for the magnetic flux.  Recall that $\mathsf{s}^\pm \sim e^{\pm i A}$ represents the gauge magnetic vector potential in this formalism, and these fields appear {\em only} in the kinetic terms of the spinon variables.  Hence in this problem the magnetic stiffness arises only dynamically: spinons propagate coherently most efficiently through a background of zero magnetic gauge flux, and thereby have lowest kinetic energy in that situation.  It is the lowering of spinon kinetic energy that is responsible for the magnetic stiffness, and hence stabilization of Coulombic (QSL and CFM) phases at zero temperature.  At $T>0$, there are now {\em two} sources of gauge fluctuations, in contrast to the situation in the pure gauge theory.  Gapped magnetic monopole excitations still exist (in these phases), but in addition we may have thermal activation of excited spinon states.  The former process is similar to that in the pure gauge theory, and is missed by gMFT.  The latter process is captured by gMFT, and acts to reduce the {\em  microscopic} magnetic stiffness even on short length scales. As this stiffness is reduced, electric field fluctuations grow in response, further decreasing the stiffness leading to a rapid explosion of gauge fluctuations and rapid reduction of the spinon bandwidth.  Once it reaches the thermal energy $k_B T$, the Coulomb phase collapses entirely as there is no microscopic stiffness to support it.  

In reality, both this process and the one due to thermally excited monopoles should be responsible for destruction of the Coulomb phase.  Since gMFT captures one but not both of these mechanisms, we expect it to be a better approximation here than in the pure gauge theory, but still susceptible to possible $O(1)$ errors.  Our expectation is that, like most mean field theories, the neglect of correlated fluctuations will lead to a reduction of the true critical temperature in comparison to the mean field result, but likely not suppress the transition entirely.  Some further evidence for this conclusion comes from examining the phase diagram more broadly.  

First, let us consider the role of symmetry.  The above discussion, and comparison to the situation in the pure gauge theory, applies best when the ground state is in the QSL phase, which breaks no symmetries and at $T>0$ can be adiabatically connected to the paramagnetic state.  In the other phases, which break symmetries, a phase transition is required at $T>0$.  In crude but physical terms, we can imagine two possible scenarios.  On one hand, confinement may occur at a lower temperature than the restoration of symmetry, so that a conventional Landau picture describes the symmetry breaking transition.  On the other hand, confinement may occur simultaneously with symmetry restoration, which is what occurs in gMFT.  In the former case, we should expect that the phase transition should be described approximately by the usual Curie-Weiss mean field theory (CWMFT) in terms of self-consistent exchange fields and spin expectation values\cite{jointpaper}, since once the transition temperature is reached, confinement is occuring on short length scales, and the microscopic spin variables are good order parameters.  It is interesting to compare the CWMFT temperature to that predicted by gMFT.  We find that the $T_c$ from CWMFT is systematically significantly larger than that found in gMFT.  For example, for the parameters corresponding to \yto, the critical temperature in CWMFT is $T_c^{\rm CWMFT} = 3.2$~K,\cite{jointpaper} while in gMFT it is $T_c^{\rm gMFT} = 0.56$~K.  The much smaller value of the critical temperature in gMFT is strong evidence that confinement physics plays a role in the transition.  Note also that the gMFT value is much closer to the observed one in \yto, supporting this notion for experiment as well.

Second, we may consider the role of Higgs condensation.  When $J_\pm$ and/or $J_{z\pm}$ are not too small, the ground state is not a deconfined but a Higgs phase, with a spinon condensate.  Consequently, the magnetic stiffness is enhanced beyond the usual dielectric form to a Meissner one, such that magnetic gauge flux is actually expelled from the system.  In particular, in these Higgs phase (the AFM and FM in Fig.~\ref{fig:MF-3dphasediag}), due to this Meissner effect, the energy of a magnetic monopole is no longer finite but actually {\em infinite}.  Consequently, magnetic monopoles cannot be thermally activated in these phases.  Therefore only the mechanism of spinon fluctuations (weakening of the Higgs condensate), which is captured by gMFT, is present in these regimes, and we expect the accuracy of gMFT to increase.  The fact that a single $T>0$ confinement-like transition appears here and is smoothly connected to the one appearing for small $J_\pm$, $J_{z\pm}$ couplings suggests that gMFT is qualitatively correct throughout the phase space.  

Based on the above reasoning, we conclude that the $T>0$ transitions for quantum spin ice in the corner of phase space studied here are qualitatively correctly described by the gMFT treatment, and should be thought of as confinement or quantum-to-classical transitions.   As a consequence of the latter, the experimental signatures (such as ``pinch points'' -- see Sec.~\ref{sec:connection-with-real}) of classical spin ice are expected in the TSL regime above the critical point.  In particular, the famous ``pinch points'' should appear in this phase.

\section{Discussion}
\label{sec:discussion}

\subsection{Connection with real materials: the case of Yb$_2$Ti$_2$O$_7$} 
\label{sec:connection-with-real}

The Hamiltonian parameters $J_{zz}$, $J_\pm$, $J_{z\pm}$ and $J_{\pm\pm}$ (see Eq.~\eqref{eq:1}) of Yb$_2$Ti$_2$O$_7$ were extracted in Ref.~\onlinecite{jointpaper}, by fitting linear spin wave theory to high-resolution inelastic neutron scattering in high field.  The accuracy of the values $J_{zz}=0.17$, $J_{\pm}=0.05$, $J_{z\pm}=-0.14$ and $J_{\pm\pm}=0.05$ meV was subsequently confirmed in Refs.~\onlinecite{applegate2012,hayre2012} through comparison of high-temperature specific heat and entropy data (for various exchange parameters reported in the literature\cite{thompson2011,jointpaper,chang2012}).    Despite the evident complete quantitative knowledge of its Hamiltonian, the nature of the low-temperature phase of Yb$_2$Ti$_2$O$_7$ in zero field is still open to debate.   Several studies find no sign of order down to the lowest accessible temperatures ($30$~mK in Ref.~\onlinecite{ross2009}),\cite{blote1969,hodges2002,gardner2004,cao2009aniso,ross2009} and diffuse neutron scattering at $T=30$~mK and $H=0$ compatible with a two-spinon continuum.\cite{jointpaper,theorypaper}  Two other neutron scattering studies have reported the presence of an ordered ferromagnetic moment.\cite{yasui2003,chang2012}   Specific heat measurements reveal strong sample dependence, which has recently been associated to Yb substitution (``stuffing'') on the Ti site,\cite{ross2012} so it is possible that such disorder modifies the zero field ground state in some samples.  However, even this is not clear.   Not knowing for sure what the low-temperature phase is, the nature of the transition observed at $T\sim200$~mK\cite{blote1969,ross2009,chang2012} also remains equivocal.

Recently, experimental\cite{chang2012} and theoretical\cite{applegate2012,hayre2012} works have also addressed the nature of the phase transition.   The authors of Ref.~\onlinecite{chang2012} argue that their experiments imply an {\em  ordered} ferromagnetic phase, and provide evidence for the first order nature of the thermal transition to this phase.  In Ref.~\onlinecite{applegate2012}, a theoretical model with third neighbor exchange (which can be considered a perturbative approximation to the full $H$\cite{jointpaper}) and consequently a ferromagnetic ground state is studied by Monte Carlo simulations, finding a first-order transition.  In our gMFT calculations, as discussed above, all the transitions with increasing $T$ are first order as well.  

Ref.~\onlinecite{chang2012} suggests that the thermal transition in \yto\ may be regarded as a ``Higgs transition''.  We would like to discuss this interpretation, in light of the one we have offered above.  The term Higgs transition has an accepted meaning in quantum field theory, where it refers to a transition which may be described as the condensation of a bosonic field carrying a non-zero gauge charge, and coupled to a dynamical gauge field.  In our formulation, such a bosonic field is the spinon, which carries the electric gauge charge, and phases with non-zero spinon condensates are indeed Higgs phases, and correspondingly have magnetic order.  The zero temperature quantum phase transitions from the CFM and QSL phases into FM and AFM phases are indeed (quantum) Higgs transitions in this sense.  In a strict sense, the situation at $T>0$ prohibits any true Higgs transitions, since the Coulomb phase itself is not sharply defined at $T>0$, i.e. there are no critical gauge fields anywhere in the phase diagram at $T>0$ since the $U(1)$ gauge fields are compact.  Even if we look for a non-strict interpretation, since the CFM and QSL phases are deconfined, transitions from them to the confined paramagnetic phase with increasing temperature can clearly not be regarded as Higgs transitions, as none of these phases have Higgs condensates.  A non-strict view of the thermal transition from the FM or AFM phases to the paramagnetic one as a Higgs transition might be possible.  However, since the FM and AFM phases are already confined states even at $T=0$, it seems unreasonable to consider spinon condensation as the mechanism for this transition.   

From our point of view, the essence of this transition is not Higgs condensation but confinement, as discussed extensively in the previous section.  According to this picture, supported by gMFT, the phase above the transition should be regarded as a classical thermal spin liquid (TSL), similar to low temperature regime of classical spin ice, and the transition may be regarded as describing the release of entropy associated with an abrupt loss of quantum coherent spin dynamics.  We would therefore expect significant differences in the inelastic spin correlations below and above the transition.  Specifically, above the transition diffuse scattering with visible pinch-point-like features would be expected, provided $T_c \ll J_{zz}$, so that the classical ice rules are not strongly violated.  Below the transition, quantum coherence is known to wash out the point points, even in the QSL regime.  Indeed, features reminiscent of pinch points were observed above $T_c$ in Ref.~\onlinecite{chang2012}\ in \yto.  Another piece of support for this interpretation is the rough agreement of the critical temperature predicted by gMFT for this confinement transition,   $T_{c}^{{\rm gMFT}}\approx560$~mK, with the experimental value $T_c^{\rm exp}\approx265$~mK\cite{ross2009}.  The agreement is much better than in the usual Curie-Weiss MFT for which $T_c^{\rm CWMFT}=3.2$~K,\cite{jointpaper} as noted above.

A more tricky issue is the presence or absence of magnetic order below $T_c$, which is controversial experimentally, and is likely to be related to the Yb stuffing mentioned earlier.  It is not unreasonable to expect the extra Yb spins to affect the state and dynamics of the system. Yet, if so, it is surprising that the microscopic model which neglects them appears to provide an excellent quantitative description of a comprehensive set of data in applied magnetic fields and/or higher temperatures.  A possible interpretation is that the ``interstitial'' spins become polarized in modest applied fields, and fluctuate paramagnetically at higher temperatures, in either case obviating their effects on the Yb sublattice spins.  At low temperature and low fields, however, the interstitial spins may be free to couple to the main sublattice, and act to effectively modify the exchange parameters, moving the material around in the general phase diagram, perhaps even into the QSL regime.  From Fig.~\ref{fig:MF-3dphasediag}, one sees that although the ground state is quite sensitive to values of the exchange parameters, the thermal transition remains confinement-like regardless of these values, consistent with observations.  Obviously this interpretation is highly speculative.  Disorder might have many other unanticipated effects, and understanding the presence or absence of magnetic order and the sample dependence of various experiments requires significant more experimental and theoretical study.  

\subsection{Conclusions}
\label{sec:conclusions}

We have studied the development of a quantum spin liquid ground state and its neighboring phases on cooling from high temperature.  We argued that in the case of quantum spin ice, quantum coherence of spins onsets in an abrupt first order transition.  When the ground state is a deconfined spin liquid, this transition may be viewed as describing the confinement of fractional spinon excitations.  Above the transition, the system has substantial entropy and behaves as a thermal spin liquid, with many of the characteristics of classical spin ice.  

In quantum pyrochlore magnets, more experimental studies on the spin correlations at intermediate temperatures would be welcome, not only from neutron scattering, but also other probes of spin dynamics such as muon spin resonance.  A thorough study of the development of the phase transition in \yto\ with applied magnetic fields would also provide considerable fuel for future theoretical work.  The techniques of the present paper, along with other approaches, can certainly address such problems.  

This raises the more general question of the existence of first order quantum to classical transitions for other models and materials with spin liquid ground states.  Most of the techniques used to study quantum spin liquids address either ground state properties (e.g. Gutzwiller variational wavefunctions, Density Matrix Renormalization Group) or are limited to relatively high temperatures (e.g. high temperature series expansion, quantum Monte Carlo), leaving the intermediate temperature regime where such a transition might occur relatively unstudied theoretically.  It may consequently be interesting to develop theoretical methods for such temperatures in the future for frustrated quantum spin models.

\begin{acknowledgments}
We thank Kate Ross, Bruce Gaulin, SungBin Lee and Shigeki Onoda for discussions.  L.B. and L.S. were supported by the DOE through Basic Energy Sciences Grant DE-FG02-08ER46524, and benefitted from the facilities of the KITP through NSF Grant PHY11-25915.
\end{acknowledgments}


\appendix

\section{Notations}
\label{sec:notations}

The local cubic bases in which the Hamiltonian Eq.~\eqref{eq:1} is expressed are the following $(\mathbf{\hat{a}}_i,\mathbf{\hat{b}}_i,\mathbf{\hat{e}}_i)$ bases 
\begin{equation}
\left\{\begin{array}{l}
\mathbf{\hat{e}}_0=(1,1,1)/\sqrt{3}\\
\mathbf{\hat{e}}_1=(1,-1,-1)/\sqrt{3}\\
\mathbf{\hat{e}}_2=(-1,1,-1)/\sqrt{3}\\
\mathbf{\hat{e}}_3=(-1,-1,1)/\sqrt{3},
\end{array}\right.,
\quad
\left\{\begin{array}{l}
\mathbf{\hat{a}}_0=(-2,1,1)/\sqrt{6}\\
\mathbf{\hat{a}}_1=(-2,-1,-1)/\sqrt{6}\\
\mathbf{\hat{a}}_2=(2,1,-1)/\sqrt{6}\\
\mathbf{\hat{a}}_3=(2,-1,1)/\sqrt{6}
\end{array}\right.,
\end{equation}
$\mathbf{\hat{b}}_i=\mathbf{\hat{e}}_i\times\mathbf{\hat{a}}_i$, such that spin $\mathbf{S}_i$ on sublattice $i$ is $\mathbf{S}_i=\mathsf{S}^+_i(\mathbf{\hat{a}}_i-i\mathbf{\hat{b}}_i)/2+\mathsf{S}^-_i(\mathbf{\hat{a}}_i+i\mathbf{\hat{b}}_i)/2+\mathsf{S}^z_i\mathbf{\hat{e}}_i$.

The $4\times4$ matrix $\gamma$ introduced in Eq.~\eqref{eq:1} is 
\begin{equation}
\gamma=\begin{pmatrix}
0 & 1 & w & w^2\\
1 & 0 & w^2 & w\\
w & w^2 & 0 & 1\\
w^2 & w & 1 & 0
\end{pmatrix},
\end{equation}
where $w=e^{2\pi i/3}$ is a third root of unity.


\section{Details of the calculations}
\label{sec:details}

Here we describe the calculations leading to Eq.~\eqref{eq:freeen} in great detail, and proceed making simplifying assumption as we go.

The decoupled Hamiltonians $H^0_\Phi$ and $H_\mathsf{s}^0$ such that the trial Hamiltonian is $H^0=H^0_\Phi+H_\mathsf{s}^0$ are given in Eqs.~\eqref{eq:h0phi} and \eqref{eq:h0s}. From them, we need to determine the free energy $F_0=-\frac{1}{\beta}\ln Z_0=F_0^\Phi+F_0^\mathsf{s}$.  The free spin part is trivial, it is the free energy of free spins in a field $\vec{\mathsf{h}}$:
\begin{eqnarray}
\label{eq:frees}
F^0_\mathsf{s}&=&-\frac{1}{\beta}\sum_{\mathbf{r}\in{\rm I},\mu}\ln\left[2\cosh\frac{\beta|\vec{\mathsf{h}}_\mu|}{2}\right]\\
&=&-\frac{1}{\beta}\sum_{\mathbf{r}\in{\rm I},\mu}\left[\ln2-\frac{1}{2}\ln\left(1-4|\langle\vec{\mathsf{s}}_{\mathbf{r},\mathbf{r}+\mathbf{e}_\mu}\rangle|^2\right)\right].\nonumber
\end{eqnarray}
To compute $F_\Phi^0$, we rewrite the first part of $H_\Phi^0$ as\cite{theorypaper}
\begin{equation}
\label{eq:Qtransform}
\sum_{\mathbf{r} \in {\rm I},{\rm II}} \frac{J}{2} Q_\mathbf{r}^2\rightarrow\sum_{\mathbf{r} \in {\rm I},{\rm II}}\left\{ \frac{J}{2} \Pi^\dagger_\mathbf{r}\Pi_\mathbf{r}+\lambda\left(\Phi_\mathbf{r}^\dagger\Phi_\mathbf{r}-1\right)\right\},
\end{equation}
where the second part of the right-hand-side is introduced to implement (a relaxed version of) the constraint $\Phi_\mathbf{r}^\dagger\Phi_\mathbf{r}=1$ (see Ref.~\onlinecite{theorypaper}). Therefore, $\lambda\in\mathbb{R}^+$ serves as a Lagrange multiplier.  We get 
\begin{widetext}
\begin{eqnarray}
  H_\Phi^0   &\rightarrow&  \sum_{\mathbf{r} \in {\rm I},{\rm II}}\left\{ \frac{J}{2} \Pi^\dagger_\mathbf{r}\Pi_\mathbf{r}+\lambda\left(\Phi_\mathbf{r}^\dagger\Phi_\mathbf{r}-1\right)\right\}   - \left\{\sum_{\mathbf{r}\in {\rm I}}  \sum_{\mu,\nu\neq\mu} t'_{\mu\nu}\, \Phi_{\mathbf{r}+\mathbf{e}_\mu}^\dagger
      \Phi_{\mathbf{r}+\mathbf{e}_\nu}^{\vphantom\dagger}
      +
       \sum_{\mathbf{r}\in
        {\rm II}}  \sum_{\mu,\nu \neq \mu} {t'_{\mu\nu}}^*\, \Phi_{\mathbf{r}-\mathbf{e}_\mu}^\dagger \Phi_{\mathbf{r}-\mathbf{e}_\nu}^{\vphantom\dagger} 
    \right\} \\
  && - \left\{ \sum_{\mathbf{r}\in {\rm I}} \sum_{\mu} \left(
    t_\mu\,\Phi_\mathbf{r}^\dagger\, \Phi_{\mathbf{r}+\mathbf{e}_\mu}^{\vphantom\dagger}
    + {\rm h.c.}\right) 
+ 
  \sum_{\mathbf{r}\in {\rm II}} \sum_{\mu} \left(
    t_\mu\,  \Phi_{\mathbf{r}-\mathbf{e}_\mu}^\dagger \Phi_{\mathbf{r}}^{\vphantom\dagger}
    + {\rm h.c.}\right) \right\},
\nonumber
\end{eqnarray}
\end{widetext}
Then, using the following Fourier transformation conventions
\begin{eqnarray}
&&x_\mathbf{k}=\frac{1}{N_{u.c.}}\sum_i x_i e^{-i\mathbf{k}\cdot\mathbf{r}_i},\quad x_\mathbf{k}^\dagger=\frac{1}{N_{u.c.}}\sum_i x_i^\dagger e^{i\mathbf{k}\cdot\mathbf{r}_i},\nonumber\\
&&\qquad x_i=\sum_\mathbf{k} x_\mathbf{k} e^{i\mathbf{k}\cdot\mathbf{r}_i},\quad x_i^\dagger=\sum_\mathbf{k} x_\mathbf{k}^\dagger e^{-i\mathbf{k}\cdot\mathbf{r}_i},
\end{eqnarray}
we arrive at
\begin{equation}
F_0^{\Phi}=\sum_\mathbf{k}\sum_{i=\pm}\omega_{\mathbf{k}}^i-2T\sum_{i=\pm}\sum_\mathbf{k}\ln\frac{1}{1-e^{-\beta\omega_{\mathbf{k}}^i}}-2\lambda N_{u.c.},
\label{eq:freephi}
\end{equation}
where $\omega_\mathbf{k}^\pm$ involve $t$ and $t'$ introduced in Eq.~\eqref{eq:h0phi}, and $J$, and where the last term comes from the constant term in Eq.~\eqref{eq:Qtransform}. Note that $\lambda$ and $\omega_\mathbf{k}^\pm$ are quantities defined from the fiducial system with Hamiltonian $H_0$ (in Ref.\onlinecite{theorypaper} their ``equivalents'' were defined from the gMFT Hamiltonian).

Let us now calculate $\langle H\rangle_0$ and $\langle H_0\rangle_0$. Going to the action formalism, with the conventions 
\begin{eqnarray}
&& x_{\Omega_n}=\frac{1}{2\pi}\int_{-\beta/2}^{+\beta/2}d\tau\, x_\tau \,e^{i\Omega_n\tau},\\
&& x_{\Omega_n}^\dagger=\frac{1}{2\pi}\int_{-\beta/2}^{+\beta/2}d\tau\,  x_\tau^\dagger \,e^{-i\Omega_n\tau},\\
&& x_{\tau}=\frac{1}{\beta}\sum_{n\in\mathbb{Z}} x_n \,e^{-i\Omega_n\tau},\\
&&x_{\tau}^\dagger=\frac{1}{\beta}\sum_{n\in\mathbb{Z}} x_n^\dagger \,e^{i\Omega_n\tau},
\end{eqnarray}
where $\Omega_n=\frac{2\pi n}{\beta}$ is the bosonic Matsubara frequency,
we find the Green's function of $H_0^\Phi$:
\begin{equation}
\label{eq:G-1}
[G^{-1}_0]=\begin{pmatrix}
\frac{1}{2J}\omega_n^2+\lambda-\tilde{L}_\mathbf{k} & -\tilde{M}_\mathbf{k} \\
-{\tilde{M}_\mathbf{k}}^* & \frac{1}{2J}\omega_n^2+\lambda-\tilde{L}_\mathbf{k}
\end{pmatrix},
\end{equation}
where 
\begin{eqnarray}
\tilde{L}_\mathbf{k}&=&\frac{1}{2}\sum_{\mu,\nu\neq\mu}\left[t'_{\mu\nu}e^{i\mathbf{k}\cdot(\mathbf{e}_\nu-\mathbf{e}_\mu)}+{t'_{\mu\nu}}^*e^{-i\mathbf{k}\cdot(\mathbf{e}_\nu-\mathbf{e}_\mu)}\right],\nonumber\\
\tilde{M}_\mathbf{k}&=&2\sum_\mu t_\mu e^{i\mathbf{k}\cdot\mathbf{e}_\mu},
\label{eq:Mtilde-t}
\end{eqnarray}
setting $t_{\mu\nu}^*=t_{\nu\mu}$. Here
$\tilde{L}$ and $\tilde{M}$ are defined in terms of $t$ and $t'$, which
are at this stage arbitrary.  In the main text we give formulae for
these quantities which are equal to the above ones when the specific
values for $t$ and $t'$ have been taken.  For simplicity we use the same
symbols for both expressions.   From this we get the spinon dispersion relations
\begin{equation}
\label{eq:omega-Jtt'}
\omega_\mathbf{k}^\pm=\sqrt{2J}\sqrt{\lambda-\tilde{L}_\mathbf{k}\pm|\tilde{M}_\mathbf{k}|}.
\end{equation}
Note that, like $\lambda$ and $\omega$, $\tilde{L}$ and $\tilde{M}$ are defined here from the fiducial Hamiltonian. So, inverting Eq.~\eqref{eq:G-1}:
\begin{widetext}
\begin{eqnarray}
G_0(\mathbf{k},\Omega_n)&=&\frac{1}{D(\mathbf{k},\Omega_n)}\begin{pmatrix}
\frac{1}{2J}\Omega_n^2+\lambda-\tilde{L}_\mathbf{k} & \tilde{M}_\mathbf{k} \\
{\tilde{M}_\mathbf{k}}^* & \frac{1}{2J}\Omega_n^2+\lambda-\tilde{L}_\mathbf{k}
\end{pmatrix}\\
&=&\frac{1}{2}\begin{pmatrix}
\frac{1}{\frac{\Omega_n^2}{2J}+\lambda-\ell^+_\mathbf{k}}+\frac{1}{\frac{\Omega_n^2}{2J}+\lambda-\ell^-_\mathbf{k}} & \frac{M_\mathbf{k}}{|M_\mathbf{k}|}\left[\frac{-1}{\frac{\Omega_n^2}{2J}+\lambda-\ell^+_\mathbf{k}}+\frac{1}{\frac{\Omega_n^2}{2J}+\lambda-\ell^-_\mathbf{k}}\right] \\
\frac{{M_\mathbf{k}}^*}{|M_\mathbf{k}|}\left[\frac{-1}{\frac{\Omega_n^2}{2J}+\lambda-\ell^+_\mathbf{k}}+\frac{1}{\frac{\Omega_n^2}{2J}+\lambda-\ell^-_\mathbf{k}}\right] & \frac{1}{\frac{\Omega_n^2}{2J}+\lambda-\ell^+_\mathbf{k}}+\frac{1}{\frac{\Omega_n^2}{2J}+\lambda-\ell^-_\mathbf{k}}
\end{pmatrix},
\end{eqnarray}
\end{widetext}
where
\begin{eqnarray}
D(\mathbf{k},\Omega_n)&=&\left(\frac{\Omega_n^2}{2J}+\lambda-\tilde{L}_\mathbf{k}\right)^2 -|\tilde{M}_\mathbf{k}|^2\\
\ell_\mathbf{k}^\pm&=&\tilde{L}_\mathbf{k}\mp|\tilde{M}_\mathbf{k}|.
\end{eqnarray}
Carrying out the sum over all integers $n$, we get
\begin{widetext}
\begin{equation}
G_0(\mathbf{k},\tau=0)=\frac{1}{\beta}\sum_{n\in\mathbb{Z}} G_0(\mathbf{k},\Omega_n) =\frac{1}{2}\sqrt{\frac{J}{2}}
\begin{pmatrix}
    \frac{\mathcal{F}_\mathbf{k}^+}{\sqrt{\lambda - \ell_\mathbf{k}^+}} + \frac{\mathcal{F}_\mathbf{k}^-}{\sqrt{\lambda - \ell_\mathbf{k}^-}} & 
    \frac{M_\mathbf{k}}{|M_\mathbf{k}|}\left[\frac{-\mathcal{F}_\mathbf{k}^+}{\sqrt{\lambda - \ell_\mathbf{k}^+}} + \frac{\mathcal{F}_\mathbf{k}^-}{\sqrt{\lambda - \ell_\mathbf{k}^-}}\right]   \\
    \frac{M_\mathbf{k}^*}{|M_\mathbf{k}|}\left[\frac{-\mathcal{F}_\mathbf{k}^+}{\sqrt{\lambda - \ell_\mathbf{k}^+}} + \frac{\mathcal{F}_\mathbf{k}^-}{\sqrt{\lambda - \ell_\mathbf{k}^-}}\right] &  \frac{\mathcal{F}_\mathbf{k}^+}{\sqrt{\lambda - \ell_\mathbf{k}^+}} + \frac{\mathcal{F}_\mathbf{k}^-}{\sqrt{\lambda - \ell_\mathbf{k}^-}}
  \end{pmatrix},
\end{equation}
\end{widetext}
where
\begin{equation}
\mathcal{F}_\mathbf{k}^\pm=\coth\left[\beta\sqrt{\frac{J}{2}}\sqrt{\lambda-\ell^\pm_\mathbf{k}}\right].
\end{equation}
As usual, we have
\begin{equation}
\langle \Phi^\dagger_{{\rm I},\mathbf{k}}\Phi_{{\rm II},\mathbf{k}}^{\vphantom{\dagger}}\rangle_0=G^0_{{\rm II},{\rm I}}(\mathbf{k}),
\end{equation}
where $G_0(\mathbf{k})=G_0(\mathbf{k},\tau=0)$, so
\begin{widetext}
\begin{eqnarray}
\langle H^0_\Phi\rangle_0&=&\frac{J}{2}\sum_{\mathbf{r} \in {\rm I},{\rm II}}  \langle\Pi^\dagger_\mathbf{r}\Pi_\mathbf{r}\rangle_0- \left\{\sum_{\mathbf{r}\in {\rm I}}  \sum_{\mu,\nu\neq\mu} t'_{\mu\nu}\, \langle\Phi_{\mathbf{r}+\mathbf{e}_\mu}^\dagger
      \Phi_{\mathbf{r}+\mathbf{e}_\nu}^{\vphantom\dagger}\rangle_0
      +
       \sum_{\mathbf{r}\in
        {\rm II}}  \sum_{\mu,\nu \neq \mu} {t'_{\mu\nu}}^*\, \langle\Phi_{\mathbf{r}-\mathbf{e}_\mu}^\dagger \Phi_{\mathbf{r}-\mathbf{e}_\nu}^{\vphantom\dagger} \rangle_0
    \right\} \\
  && - \left\{ \sum_{\mathbf{r}\in {\rm I}} \sum_{\mu} \left(
    t_\mu\,\langle\Phi_\mathbf{r}^\dagger\, \Phi_{\mathbf{r}+\mathbf{e}_\mu}^{\vphantom\dagger}\rangle_0
    + {\rm h.c.}\right) 
+ 
  \sum_{\mathbf{r}\in {\rm II}} \sum_{\mu} \left(
    t_\mu\,  \langle\Phi_{\mathbf{r}-\mathbf{e}_\mu}^\dagger \Phi_{\mathbf{r}}^{\vphantom\dagger}\rangle_0
    + {\rm h.c.}\right) \right\}.
\nonumber
\end{eqnarray}
Now, recall:
\begin{eqnarray}
  \langle H\rangle_0 & = & \frac{J_{zz}}{2} \sum_{\mathbf{r} \in {\rm I},{\rm II}}  \langle\Pi^\dagger_\mathbf{r}\Pi_\mathbf{r}\rangle_0  - J_\pm \left\{\sum_{\mathbf{r}\in {\rm I}}  \sum_{\mu,\nu\neq\mu} \overline{\sf s}_\mu\,\overline{\sf s}_\nu^*\, \langle\Phi_{\mathbf{r}+\mathbf{e}_\mu}^\dagger
      \Phi_{\mathbf{r}+\mathbf{e}_\nu}^{\vphantom\dagger}\rangle_0
      +
       \sum_{\mathbf{r}\in
        {\rm II}}  \sum_{\mu,\nu \neq \mu} \overline{\sf s}_\mu^*\,\overline{\sf
        s}_\nu\, \langle\Phi_{\mathbf{r}-\mathbf{e}_\mu}^\dagger \Phi_{\mathbf{r}-\mathbf{e}_\nu}^{\vphantom\dagger} \rangle_0
    \right\} \\
  && - J_{z\pm} \left\{ \sum_{\mathbf{r}\in {\rm I}} \sum_{\mu,\nu\neq\mu} \left(
    \gamma^*_{\mu\nu}\, \overline{\sf s}_\mu^z\, \overline{\sf
        s}_\nu^*\,\langle\Phi_\mathbf{r}^\dagger\, \Phi_{\mathbf{r}+\mathbf{e}_\nu}^{\vphantom\dagger}\rangle_0
    + {\rm h.c.}\right) 
+ 
  \sum_{\mathbf{r}\in {\rm II}} \sum_{\mu,\nu\neq\mu} \left(
    \gamma^*_{\mu\nu}\, \overline{\sf s}_\mu^z\, \overline{\sf
        s}_\nu^*\,  \langle\Phi_{\mathbf{r}-\mathbf{e}_\nu}^\dagger \Phi_{\mathbf{r}}^{\vphantom\dagger}\rangle_0
    + {\rm h.c.}\right) \right\},
\nonumber
\end{eqnarray}
where we have defined $\overline{x}=\langle x\rangle_0$, and with
\begin{equation}
\langle \Phi_{\mathbf{r}+\mathbf{e}_\mu}^\dagger\Phi_{\mathbf{r}+\mathbf{e}_\nu}^{\vphantom{\dagger}}\rangle_0=
\langle \Phi_{\mathbf{e}_\mu-\mathbf{e}_\nu}^\dagger\Phi_{\mathbf{0}}^{\vphantom{\dagger}}\rangle_0=\sum_\mathbf{k}\langle\Phi_{{\rm II},\mathbf{k}}^\dagger\Phi_{{\rm II},\mathbf{k}}^{\vphantom{\dagger}}\rangle_0 e^{i\mathbf{k}\cdot(\mathbf{e}_\mu-\mathbf{e}_\nu)}=\frac{1}{2N_{u.c.}}\sqrt{\frac{J}{2}}\sum_\mathbf{k}\left[\frac{\mathcal{F}_\mathbf{k}^+}{\sqrt{\lambda - \ell_\mathbf{k}^+}} + \frac{\mathcal{F}_\mathbf{k}^-}{\sqrt{\lambda - \ell_\mathbf{k}^-}}\right]e^{i\mathbf{k}\cdot(\mathbf{e}_\mu-\mathbf{e}_\nu)}
\label{eq:PhiexpectNNN}
\end{equation}
and
\begin{equation}
\label{eq:PhiexpectNN}
\langle \Phi_{\mathbf{r}}^\dagger\Phi_{\mathbf{r}+\mathbf{e}_\nu}^{\vphantom{\dagger}}\rangle_0
=\sum_\mathbf{k}\langle\Phi_{{\rm I},\mathbf{k}}^\dagger\Phi_{{\rm II},\mathbf{k}}^{\vphantom{\dagger}}\rangle_0 e^{-i\mathbf{k}\cdot\mathbf{e}_\nu}=-\frac{1}{2N_{u.c.}}\sqrt{\frac{J}{2}}\sum_\mathbf{k}\frac{\tilde{M}_\mathbf{k}^*}{|\tilde{M}_\mathbf{k}|}\left[\frac{\mathcal{F}_\mathbf{k}^+}{\sqrt{\lambda - \ell_\mathbf{k}^+}} + \frac{-\mathcal{F}_\mathbf{k}^-}{\sqrt{\lambda - \ell_\mathbf{k}^-}}\right]e^{-i\mathbf{k}\cdot\mathbf{e}_\nu}.
\end{equation}
Finally,
\begin{eqnarray}
\label{eq:20}
\langle H - H_0\rangle_0&=&\langle H\rangle_0-\langle H_0^\Phi\rangle_0-\langle H_0^\mathsf{s}\rangle_0\\
&=&
\frac{J_{zz}-J}{2} \sum_{\mathbf{r} \in {\rm I},{\rm II}}  \langle\Pi^\dagger_\mathbf{r}\Pi_\mathbf{r}\rangle_0 \\
&&  - \left\{\sum_{\mathbf{r}\in {\rm I}}  \sum_{\mu,\nu\neq\mu} \left(J_\pm \overline{\sf s}_\mu\,\overline{\sf s}_\nu^*-t'_{\mu\nu}\right) \langle\Phi_{\mathbf{r}+\mathbf{e}_\mu}^\dagger
      \Phi_{\mathbf{r}+\mathbf{e}_\nu}^{\vphantom\dagger}\rangle_0
      +
       \sum_{\mathbf{r}\in
        {\rm II}}  \sum_{\mu,\nu \neq \mu} \left(J_\pm \overline{\sf s}_\mu^*\,\overline{\sf
        s}_\nu - {t_{\mu\nu}'}^*\right) \langle\Phi_{\mathbf{r}-\mathbf{e}_\mu}^\dagger \Phi_{\mathbf{r}-\mathbf{e}_\nu}^{\vphantom\dagger} \rangle_0
    \right\} \nonumber \\
  && -\left\{ \sum_{\mathbf{r}\in {\rm I}} \sum_{\mu,\nu} \left[
     \left(J_{z\pm} \gamma^*_{\mu\nu}\, \overline{\sf s}_\mu^z\, \overline{\sf
        s}_\nu^*-\frac{t_\nu}{4}\right)\langle\Phi_\mathbf{r}^\dagger\, \Phi_{\mathbf{r}+\mathbf{e}_\nu}^{\vphantom\dagger}\rangle_0
    + {\rm h.c.}\right]\right. \\
&&\left.\qquad\qquad\qquad\qquad+ 
  \sum_{\mathbf{r}\in {\rm II}} \sum_{\mu,\nu} \left[
     \left(J_{z\pm} \gamma^*_{\mu\nu}\, \overline{\sf s}_\mu^z\, \overline{\sf s}_\nu^* -\frac{t_\nu}{4}\right)  \langle\Phi_{\mathbf{r}-\mathbf{e}_\nu}^\dagger \Phi_{\mathbf{r}}^{\vphantom\dagger}\rangle_0
    + {\rm h.c.}\right] \right\}\nonumber\\
&& +\sum_{\mathbf{r}\in{\rm I}}\sum_\mu \vec{\mathsf{h}}_\mu(\mathbf{r})\cdot\langle\vec{\mathsf{s}}_{\mathbf{r},\mathbf{r}+\mathbf{e}_\mu}\rangle_0.\nonumber
\end{eqnarray}
\end{widetext}
Now, simply assuming $|\langle\vec{\mathsf{s}}_{\mathbf{r},\mathbf{r}+\mathbf{e}_\mu}\rangle_0|=\mathsf{s}$ to be independent of $\mu$, we have already:
\begin{eqnarray}
\langle H_\mathsf{s}^0\rangle_0&=&-\sum_{\mathbf{r}\in{\rm I}}\sum_\mu \vec{\mathsf{h}}_\mu(\mathbf{r})\cdot\langle\vec{\mathsf{s}}_{\mathbf{r},\mathbf{r}+\mathbf{e}_\mu}\rangle_0\\
&=&-\frac{4N_{u.c.}\mathsf{s}}{\beta}\ln\frac{1+2\mathsf{s}}{1-2\mathsf{s}},
\label{eq:hsexpect}
\end{eqnarray}
and if we use the Ansatz Eq.~\eqref{eq:ansatz},
\begin{equation}
\overline{\mathsf{s}}^z_\mu=\mathsf{s}\,\varepsilon_\mu\sin\theta\qquad\mbox{and}\qquad\overline{\mathsf{s}}^-_\mu=\mathsf{s}\,\cos\theta,\nonumber
\end{equation}
then 
\begin{eqnarray}
&&\sum_\mu J_{z\pm}\gamma_{\mu\nu}^*\overline{\mathsf{s}}_\mu^z\overline{\mathsf{s}}_\nu^*=2\mathsf{s}^2J_{z\pm}\varepsilon_\nu\cos\theta\sin\theta=\mathsf{s}^2J_{z\pm}\varepsilon_\nu\sin2\theta,\nonumber\\
&&\quad\mbox{and}\quad J_\pm \overline{\sf s}_\mu\,\overline{\sf s}_\nu^*=J_\pm\mathsf{s}^2\cos^2\theta.
\end{eqnarray}
We now take
\begin{equation}
t'_{\mu\nu}=t'\qquad\mbox{and}\qquad t_\nu = \varepsilon_\nu t,
\end{equation}
so that Eq.~\eqref{eq:20} becomes
\begin{widetext}
\begin{eqnarray}
\label{eq:26}
\langle H - H_0\rangle_0&=&
\frac{J_{zz}-J}{2} \sum_{\mathbf{r} \in {\rm I},{\rm II}}  \langle\Pi^\dagger_\mathbf{r}\Pi_\mathbf{r}\rangle_0 \\
&&  - \left(\frac{J_1}{2}-t'\right)\left\{\sum_{\mathbf{r}\in {\rm I}}  \sum_{\mu,\nu\neq\mu}  \langle\Phi_{\mathbf{r}+\mathbf{e}_\mu}^\dagger
      \Phi_{\mathbf{r}+\mathbf{e}_\nu}^{\vphantom\dagger}\rangle_0
      +
       \sum_{\mathbf{r}\in
        {\rm II}}  \sum_{\mu,\nu \neq \mu} \langle\Phi_{\mathbf{r}-\mathbf{e}_\mu}^\dagger \Phi_{\mathbf{r}-\mathbf{e}_\nu}^{\vphantom\dagger} \rangle_0
    \right\} \nonumber \\
  && -\left(\frac{J_2}{2}-t\right)
        \left\{ \sum_{\mathbf{r}\in {\rm I}} \sum_{\nu} \left[\varepsilon_\nu
     \langle\Phi_\mathbf{r}^\dagger\, \Phi_{\mathbf{r}+\mathbf{e}_\nu}^{\vphantom\dagger}\rangle_0
    + {\rm h.c.}\right] 
+ 
  \sum_{\mathbf{r}\in {\rm II}} \sum_{\nu} \left[\varepsilon_\nu
      \langle\Phi_{\mathbf{r}-\mathbf{e}_\nu}^\dagger \Phi_{\mathbf{r}}^{\vphantom\dagger}\rangle_0
    + {\rm h.c.}\right] \right\}\nonumber\\
&& +\sum_{\mathbf{r}\in{\rm I}}\sum_\mu \vec{\mathsf{h}}_\mu(\mathbf{r})\cdot\langle\vec{\mathsf{s}}_{\mathbf{r},\mathbf{r}+\mathbf{e}_\mu}\rangle_0,\nonumber
\end{eqnarray}
\end{widetext}
where
\begin{equation}
J_1=2J_\pm\mathsf{s}^2\cos^2\theta\qquad\mbox{and}\qquad J_2=2J_{z\pm}\mathsf{s}^2\sin2\theta.
\end{equation}
Finally, setting $J=J_{zz}$, $t'_{\mu\nu}=t'=\frac{J_1}{2}=J_\pm\mathsf{s}^2\cos^2\theta$ and $t_\mu=\varepsilon_\mu t=\varepsilon_\mu \frac{J_2}{2}=\varepsilon_\mu J_{z\pm}\mathsf{s}^2\sin2\theta$, and using Eqs.~\eqref{eq:PhiexpectNNN}, \eqref{eq:PhiexpectNN} and \eqref{eq:hsexpect} we recover Eq.~\eqref{eq:freeen}.


\section{Explicit expression of the $I_3=1$ constraint in the condensed and deconfined phases}
\label{sec:I3}

The constraint on the spinons (rotor operators) $\Phi_\mathbf{r}^\dagger\Phi^{\vphantom{\dagger}}_\mathbf{r}=1$ is enforced in the form $\langle\Phi_\mathbf{r}^\dagger\Phi^{\vphantom{\dagger}}_\mathbf{r}\rangle=1$, i.e.
\begin{equation}
1=I_3=\frac{1}{2N_{u.c.}}\sqrt{\frac{J_{zz}}{2}}\sum_\mathbf{k}\left[\frac{\mathcal{F}_\mathbf{k}^+}{\sqrt{\lambda-\ell_\mathbf{k}^+}}+\frac{\mathcal{F}_\mathbf{k}^-}{\sqrt{\lambda-\ell_\mathbf{k}^-}}\right],
\label{eq:I3append}
\end{equation}
where
\begin{equation}
\mathcal{F}^\pm_\mathbf{k}=\coth\left[\beta\sqrt{\frac{J_{zz}}{2}}\sqrt{\lambda-\ell_\mathbf{k}^\pm}\right],\qquad\ell^\pm_\mathbf{k}=\tilde{L}_\mathbf{k}\mp|\tilde{M}_\mathbf{k}|.
\end{equation}

While in the deconfined phases the sum in Eqs.~\eqref{eq:I3} and \eqref{eq:I3append} can be turned simply into an integral (since $\lambda-\ell^\pm_{\mathbf{k}}>0$ for all $\mathbf{k}$), in the condensed phases, the spinon dispersion relation hits zero at a wavevector $\mathbf{k}_0$, i.e. such that  $\ell^-_{\mathbf{k}_0}=\lambda_{\min}=\max_\mathbf{k}\ell^-_\mathbf{k}$, and one should allow for a subextensive part in $\lambda$. We write $\lambda=\lambda_{\rm min}+\hat{\delta}T/N_{u.c.}$, where $\hat{\delta}=O(1)$, with $\hat{\delta}$ positive and independent of $T$. This leads to (see also the Supplemental Material of Ref.~\onlinecite{theorypaper}):
\begin{equation}
I_3=I_3^{\rm min}+I_3',
\end{equation}
where $I_3'$ is the right-hand-side of Eq.~\eqref{eq:I3append} turned into an integral and evaluated at $\lambda=\lambda_{\rm min}$, i.e.
\begin{equation}
I_3'=\frac{1}{2}\sqrt{\frac{J_{zz}}{2}}\int_\mathbf{k}\left[\frac{\mathcal{F}^+_\mathbf{k}}{\sqrt{\lambda_{\rm min}-\ell^+_\mathbf{k}}}+\frac{\mathcal{F}^-_\mathbf{k}}{\sqrt{\lambda_{\rm min}-\ell^-_\mathbf{k}}}\right],
\end{equation}
and $I_3^{\rm min}$ is the part of $\frac{1}{2N_{u.c.}}\sqrt{\frac{J_{zz}}{2}}\sum_{i=\pm1}\frac{\mathcal{F}_{\mathbf{k}_0}^i}{\sqrt{\lambda-\ell_{\mathbf{k}_0}^i}}$ which does not vanish when $N_{u.c.}\rightarrow\infty$. Defining $\rho=I_3^{\rm min}$, we find, in terms of $\hat{\delta}$,
 \begin{equation}
\rho=\begin{cases}
\frac{1}{2\hat{\delta}}&\mbox{for }\theta,\mathsf{s},J_{z\pm}\neq0\\
\frac{1}{\hat{\delta}}&\mbox{otherwise}
\end{cases},
\label{eq:rho}
\end{equation}
or more generally, considering the rotor constraint, $\rho=1-I_3'$.

Note: {\it (i)} the difference in the exponent compared with the zero
temperature case explored in Ref.~\onlinecite{theorypaper}, which comes
from the coefficient modification involved with
$\mathcal{F}^\pm_\mathbf{k}$, {\it (ii)} we used a hat on $\delta$
because its definition differed from its ``equivalent'' at $T=0$. 

A final remark is in order: in $\sum_\mathbf{k}\sum_{i=\pm1}\left[2T\ln\frac{1}{1-e^{-\beta\omega_\mathbf{k}^i}}\right]$ of Eq.~\eqref{eq:freeen}, the $\mathbf{k}_0$ term is of order $\ln N_{u.c.}$ (for $N_{u.c.}$ large at fixed $T$) in the condensed phases. Since $\ln N_{u.c.}=o(N_{u.c.})$, this contribution is actually negligible at large $N_{u.c.}$ compared with the main contribution to the free energy, which is extensive.


\section{Comparison with zero temperature}
\label{sec:zerotemp}

Here we outline the procedure described in Ref.~\onlinecite{theorypaper}, i.e. in the case of zero temperature, and show that this case is recovered when we take $T\rightarrow0$ in the present work.

\subsection{Energy at $T=0$ as derived in Savary and Balents}

In Ref.~\onlinecite{theorypaper}, we reported
\begin{equation}
\langle H\rangle_{{\rm gMFT}, T=0}=E_{\rm GS}=N_{u.c.}\left(\epsilon_{av}+\epsilon_{kin}\right),
\label{eq:zeroTvarenergy}
\end{equation}
with
\begin{eqnarray}
\label{eq:eav}
\epsilon_{av}&=&-2I_2(\theta,\lambda)\cos^2\theta J_\pm-4I_1(\theta,\lambda)\sin2\theta J_{z\pm}\\
\epsilon_{kin}&=&\frac{1}{2}\int_{\mathbf{k}}\left(\omega_\mathbf{k}^+(\theta,\lambda)+\omega_\mathbf{k}^-(\theta,\lambda)\right),
\label{eq:ekin}
\end{eqnarray}
where $\omega^\pm_\mathbf{k}=\sqrt{2J_{zz}}z^\pm_\mathbf{k}$. $\lambda$ will have been determined by solving $I_3=1$, and $\theta$ either by solving the consistency equations (and choosing the lowest energy solution), or by minimizing the Eq.~\eqref{eq:zeroTvarenergy} form of the energy.

The energy can also be derived through another procedure, which we call ``by decomposition.''  The energy found in such a way is {\em  not} variational (i.e. the ground state energy cannot be found by minimizing it over $\theta$); the ground state energy is found by plugging in the parameter values found by solving the consistency equations. For those values the decomposition and the variational forms of the energy yield equal values. The energy is found by analyzing what the decoupled problem is equivalent to.  The mean-field Hamiltonian, found using the usual decomposition (see Eq.~(6) of Ref.~\onlinecite{theorypaper}), is
\begin{equation}
H^{\rm MF}=H_{\rm spinon}^{\rm MF}+H^{\rm MF}_{\rm spin}-E^{\rm MF}_{\rm const},
\end{equation}
so that
\begin{equation}
E^{\rm MF}=N_{u.c.}\left(\epsilon_{\rm spinon}^{\rm MF}+\epsilon_{\rm spin}^{\rm MF}-\epsilon_{\rm const}^{\rm MF}\right).
\end{equation}
We find
\begin{eqnarray}
\epsilon_{\rm spinon}^{\rm MF}&=&\frac{1}{2\pi}\int_{\mathbf{k}}\left(\omega^+_\mathbf{k}+\omega^-_\mathbf{k}\right)-2\lambda\\
\epsilon_{\rm spin}^{\rm MF}&=&-2\sqrt{\overline{\mathsf{h}_{\rm MF}^z}^2+\overline{\mathsf{h}_{\rm MF}^x}^2}\\
\epsilon_{\rm const}^{\rm MF}&=&-4I_2\cos^2\theta J_\pm-8 I_1\sin2\theta J_{z\pm},
\end{eqnarray}
where $I_1$, $I_2$, and $\overline{\vec{\mathsf{h}}_{\rm MF}}$ were defined in Ref.~\onlinecite{theorypaper}. It turns out that we find, empirically (i.e. numerically), 
\begin{equation}
\forall\;\theta\qquad E_{\rm spinon}^{\rm MF}=E_{\rm GS},
\label{eq:empirical}
\end{equation}
provided $\lambda$ is chosen such that $I_3=1$ is satisfied. 

Note that, to avoid any confusion, we refrained from calling $\epsilon_{\rm spinon}^{\rm MF}$ (resp. $\epsilon_{\rm spin}^{\rm MF}$) $\epsilon_\Phi^{\rm MF}$ (resp. $\epsilon_{\mathsf{s}}^{\rm MF}$) as they do not come from  $H_\Phi^0$ and $H_\mathsf{s}^0$ of the present paper.

\subsection{$T\rightarrow0$ limit of the variational free energy}

We now go back to the variational free energy at $T>0$, Eq.~\eqref{eq:freeen}, and take its $T\rightarrow0$ limit.  Since $\lim_{T\rightarrow0}\mathsf{s}=1/2$, the first part of the first term of Eq.~\eqref{eq:freeen} vanishes in the $T\rightarrow0$ limit. Therefore,
\begin{eqnarray}
\label{eq:FvT0limit}
&&\lim_{T\rightarrow0}F_v=-2 N_{u.c.}\lim_{T\rightarrow0}\lambda\\
&&\qquad+\sum_\mathbf{k}\sum_{i=\pm1}\left\{\lim_{T\rightarrow0}\omega_\mathbf{k}^i-\lim_{T\rightarrow0}\left[2T\ln\frac{1}{1-e^{-\beta\omega_\mathbf{k}^i}}\right]\right\}.\nonumber
\end{eqnarray}
Since $\lambda$ is determined through the $I_3=1$ constraint, let us address the becoming of the latter. It is
\begin{equation}
1=\frac{1}{2N_{u.c.}}\sqrt{\frac{J_{zz}}{2}}\sum_\mathbf{k}\sum_{i=\pm1}\lim_{T\rightarrow0}\frac{\mathcal{F}^i_\mathbf{k}}{\sqrt{\lambda-\ell_\mathbf{k}^i}},
\end{equation}
with $\mathcal{F}$ defined in Eq.~\eqref{eq:Fcaldef},
\begin{equation}
\mathcal{F}^\pm_\mathbf{k}=\coth\left[\beta\sqrt{\frac{J_{zz}}{2}}\sqrt{\lambda-\ell_\mathbf{k}^\pm}\right].
\end{equation}
In the deconfined phases, since $\lambda-\ell^\pm_\mathbf{k}>0$ for all
$\mathbf{k}$, $\mathcal{F}^i_\mathbf{k}\rightarrow1$ trivially, and it
is obvious the $I_3=1$ constraint reduces to that found at zero
temperature in Ref.~\onlinecite{theorypaper} thanks to the choice
(described in Section~\ref{sec:gauge-mean-field} and
Appendix~\ref{sec:details}) of parameters $t$, $t'$ and $J$ of the
fiducial Hamiltonian. It follows immediately that
$\lim_{T\rightarrow0}\lambda^{\rm decon}=\lambda^{\rm decon}(T=0)$. In
the condensed phases, as described in Ref.~\onlinecite{theorypaper} and
Appendix~\ref{sec:I3}, the sum is better split into a $\mathbf{k}_0$
term (which we call $\rho$) for which
$\ell^-_{\mathbf{k}_0}=\lambda_{\min}=\max_\mathbf{k}\ell^-_\mathbf{k}$,
and remaining terms $I_3'$.  Defining $\lambda=\lambda_{\rm
  min}+\hat{\delta}T/N_{u.c.}$ in the condensed phases and taking the
$N_{u.c.}\rightarrow\infty$ limit before the $T\rightarrow0$ limit
(since physically we are interested in low but non-zero temperature but
thermodynamically large systems), we find
$\lim_{T\rightarrow0}\lambda_{\rm min}=\lambda_{\rm min}(T=0)$
$\lim_{T\rightarrow0}I_3'=I_3'(T=0)$, which, from $\rho=1-I_3'$, implies
$\lim_{T\rightarrow0}\rho=\rho(T=0)$, and as a consequence,
\begin{equation}
\lim_{T\rightarrow0}\lambda=\lambda(T=0),\quad\mbox{and}\quad
\lim_{T\rightarrow0}\omega^i_\mathbf{k}=\omega^i_\mathbf{k}(T=0).
\end{equation}
Finally, the very last term $\lim_{T\rightarrow0}\left[2T\ln\frac{1}{1-e^{-\beta\omega_\mathbf{k}^i}}\right]$ of Eq.~\eqref{eq:FvT0limit} goes trivially to zero in the deconfined phases. In the condensed phases, as noted in Appendix~\ref{sec:I3}, the $\mathbf{k}_0$ term goes as $\ln N_{u.c.}$ for $N_{u.c.}$ large (at fixed $T$), but the zero-temperature limit takes this term to zero so that $\lim_{T\rightarrow0}\left[2T\ln\frac{1}{1-e^{-\beta\omega_\mathbf{k}^i}}\right]=0$ in the condensed phases as well.

Finally, we arrive at
\begin{eqnarray}
\lim_{T\rightarrow0}F_v&=&-2N_{u.c.}\lambda(T=0)+\sum_{\mathbf{k}}\sum_{i=\pm1}\omega^i_\mathbf{k}(T=0)\nonumber\\
&=&E_{\rm spinon}.
\label{eq:finallimit}
\end{eqnarray}
Using the empirical evidence Eq.~\eqref{eq:empirical}, this proves that the variational form of the ground state energy is recovered when we take the zero temperature limit of our variational free energy. Note that Eq.~\eqref{eq:finallimit} could also be seen as a convoluted proof of Eq.~\eqref{eq:empirical}!

\subsection{Discrepancy with the $T=0$ phase diagram computed in Savary and Balents}

To obtain the phase diagram of Figure 1 of Ref.~\onlinecite{theorypaper}, we did not minimize the variational ground state energy at $T=0$, but rather solved the consistency equations Eq.~(11) (of Ref.~\onlinecite{theorypaper}), and selected those with lowest energy.  

\begin{figure}[htbp]
\begin{center}
\includegraphics[width=3.3in]{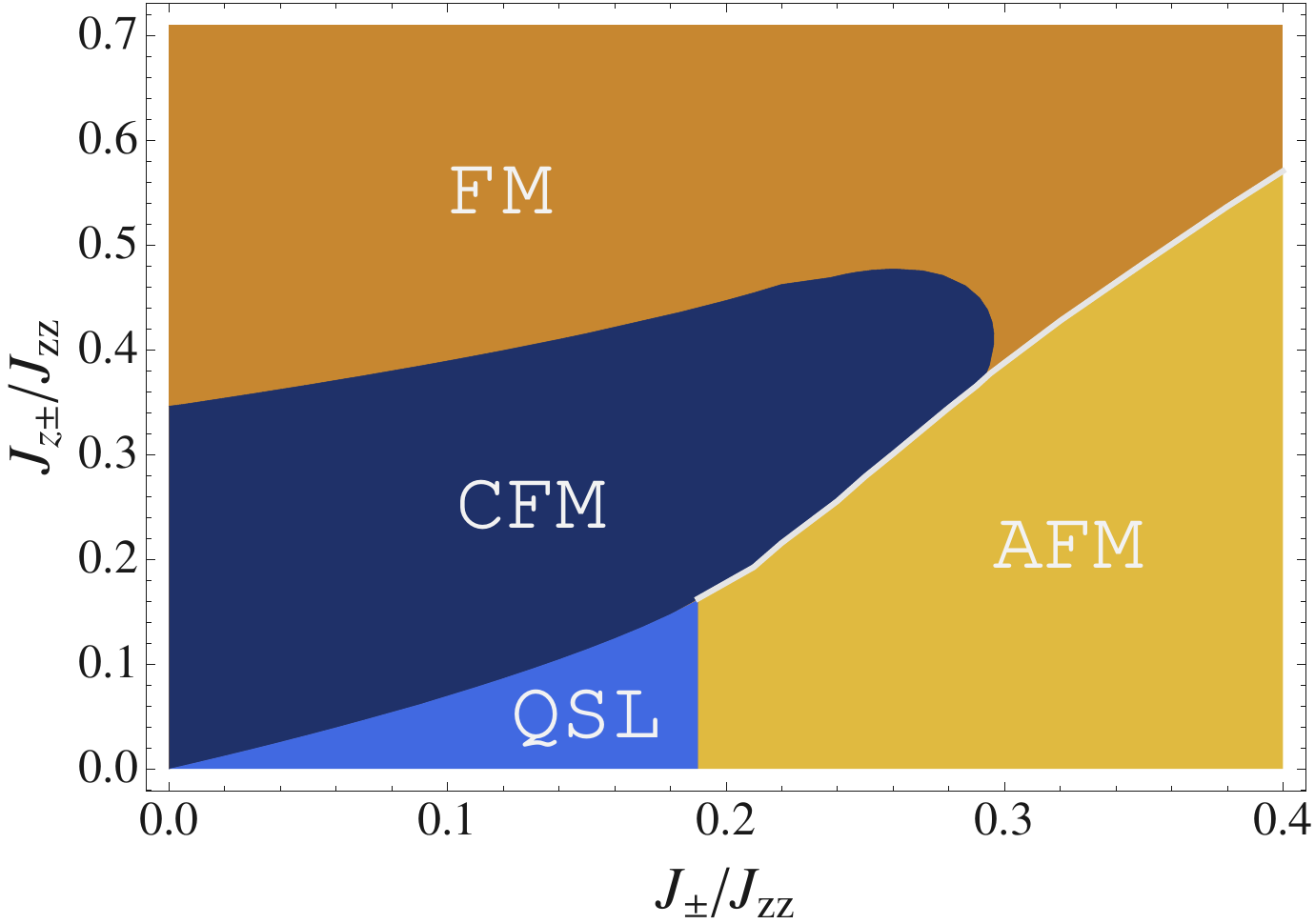}
\caption{(Color online) Zero-temperature gauge mean field phase diagram obtained for $J_{\pm\pm}=0$ and $J_{zz}>0$.  ``QSL'', ``CFM'', ``FM'', ``AFM'' denote the $U(1)$ Quantum Spin Liquid, Coulomb Ferromagnet, standard ferromagnet, and standard antiferromagnet, respectively.}
\label{fig:T=0correctdiag}
\end{center}
\end{figure}

In Ref.~\onlinecite{theorypaper}, the consistency equation Eq.~(11) as
well as the variational form of the energy Eqs.~(12) and (13)
(Eqs.~\eqref{eq:eav} and \eqref{eq:ekin} here) involve the sums $I_1$
and $I_2$ which are subject to greater numerical errors than the
variational form (of the type of $E_{\rm spinon}^{\rm MF}$) presented
here.  This led to a small mistake in the position of the AFM-FM phase
boundary in Ref.~\onlinecite{theorypaper}.  We believe the slightly
modified diagram presented (Figure~\ref{fig:T=0correctdiag}) here is correct.  An Erratum with this
correction is being simultaneously submitted to Physical Review Letters.

\section{Analytical phase transitions in the small parameter regime $J_\pm,J_{z\pm}\ll J_{zz}$}
\label{sec:smallparams}

Here we look at simple limits and find approximate phase transitions analytically, taking advantage of the fact that the phase transitions to the TSL occur at low temperature.

\subsection{$\mathsf{s}=0$}

For $\mathsf{s}=0$,
\begin{eqnarray}
F_v(\mathsf{s}=0)/N_{u.c.}&=&2\sqrt{\lambda}\left(\sqrt{2J_{zz}}-\sqrt{\lambda}\right)\\
&&-4T\left(\ln2-\ln\left[1-e^{-\frac{\sqrt{2J_{zz}\lambda}}{T}}\right]\right),\nonumber
\end{eqnarray}
and
\begin{equation}
1=\sqrt{\frac{J_{zz}}{2\lambda}}\coth\left[\frac{1}{T}\sqrt{\frac{J_{zz}\lambda}{2}}\right].
\end{equation}
For small enough temperature, we find $\lambda=J_{zz}/2$, which leads to
\begin{equation}
F_v(\mathsf{s}=0)/N_{u.c.}\approx J_{zz}-4T\ln2.
\label{eq:Fvs0}
\end{equation}

\subsection{$\mathsf{s}=1/2$}

When $\mathsf{s}=1/2$ (which encompasses but is not a priori restricted to the $T=0$ case),
\begin{eqnarray}
\label{eq:Fvs1/2}
&&F_v(\mathsf{s}=1/2)/N_{u.c.}\\
&&\qquad\quad=F_v(T=0)/N_{u.c.}-2T\sum_{i=\pm}\int_\mathbf{q}\ln\frac{1}{1-e^{-\omega_{\mathbf{q}}^i/T}},\nonumber
\end{eqnarray}
so, for small enough temperature, the variational free energy at
$T\neq0$ is almost constant and equal to that at $T=0$.  Eq.~\eqref{eq:Fvs1/2} is
valid both in the condensed and uncondensed phases.  The absence of a
distinct correction to this form in the presence of a condensate may be
interpreted physically as the fact that the condensate carries zero
entropy.  Since the final term in Eq.~\eqref{eq:Fvs1/2} is a measure of the
entropy, it is not corrected by a condensate.

\subsection{Transition to the $\mathsf{s}=0$ state (TSL) (if it indeed occurs at small $T$ and small $J_\pm$)}

We apply the results Eqs.~\eqref{eq:Fvs0} and \eqref{eq:Fvs1/2} to find analytical forms of the transitions.

\subsubsection{For $J_{z\pm}=0$}

\begin{equation}
T_c=\frac{3J^c_\pm{}^2}{16J_{zz}\ln2}\qquad\Longleftrightarrow\qquad J^c_\pm=4\,\sqrt{\frac{T_c J_{zz}\ln2}{3}}.
\end{equation}

\subsubsection{For $J_{z\pm}\neq0$}

\begin{equation}
T_c J_{zz}=\frac{3\cos^4\theta}{16\ln2}J^{c}_\pm{}^2+\frac{\sin^22\theta}{4\ln2}J^{c}_{z\pm}{}^2,
\end{equation}
where $\theta$ needs to have been well chosen to minimize the $T=0$ energy, i.e.
\begin{equation}
T_cJ_{zz}=\frac{3J^c_\pm{}^2}{16J_{zz}\ln2}\quad\mbox{for}\quad J_{z\pm}\leq\frac{\sqrt{3}}{2\sqrt{2}}J_{\pm}\quad\mbox{where}\; \theta=0,
\end{equation}
and at
\begin{eqnarray}
&&T_cJ_{zz}=\frac{1}{\ln2}\frac{4J^{c}_{z\pm}{}^4}{16{J^{c}_{z\pm}}^2-3J_\pm^c{}^2}\;\\
&&\qquad\qquad\qquad\qquad\mbox{for}\; J_{z\pm}>\frac{\sqrt{3}}{2\sqrt{2}}J_{\pm}\;\mbox{where}\; \theta\neq0.\nonumber
\end{eqnarray}


\section{Calculation of the phase diagram and representation of 3D surfaces and cuts}
\label{sec:figures-calc}

The phase diagram, Figure~\ref{fig:MF-3dphasediag} and the cuts, Figure~\ref{fig:threecuts}, was obtained by sampling points separated by $0.1$ increments in the $J_\pm/J_{zz}$ direction, $0.0125$ to up to $0.05$ in the $J_{z\pm}/J_{zz}$ direction and $0.0025$ in the $T/J_{zz}$ direction in regions surrounding a phase transition.

Minimization was obtained by comparing values of $F_v$ for values $\mathsf{s}=0.0001$, $0.005$, $0.01$, $0.02$, $0.05$, $0.1$, $0.2$, $0.3$, $0.4$, $0.45$, $0.46$, $0.47$, $0.48$, $0.49$, $0.495$, $0.496$, $0.497$, $0.498$, $0.499$, $0.49999$, and $\theta=n\pi/32$, $n=0,..,7$ for each of the sampled points.  Which phase each point belonged to was determined according to Table~\ref{tab:phase-class}.

\begin{table}[htdp]
  \caption{Criteria for determining the ground state phase. Note that, in practice, the criterion for $\theta$ or $\mathsf{s}$ zero or nonzero is $\theta$ or $\mathsf{s}$ smaller or greater than $10^{-6}$.}
\begin{center}
\begin{tabular}{|c|c|c||c|}
\hline \quad$\rho\quad$ & \quad$\theta\quad$ & \quad$ \mathsf{s}\quad$ & \quad phase$\quad$  \\ \hline 
$0$ & $0$ & $\neq0$ & QSL \\ \hline
$0$ & $\neq0$ & $\neq0$ & CFM \\ \hline
$\neq0$ & $\neq0$ & $\neq0$ & FM \\ \hline
$\neq0$ & $0$ & $\neq0$ & AFM \\ \hline
$0$ & $0$ & $0$ & TSL \\ \hline
\end{tabular}
\end{center}
\label{tab:phase-class}
\end{table}%

A set $\mathcal{E}$ of phase transition points was subsequently obtained by taking the midpoints (along well-chosen lines) between sampled points not belonging to the same phase.

The phase transition {\em surfaces} were then obtained by triangulating the projections of the points of $\mathcal{E}$ onto appropriate planes.

The triangle edges were subsequently parametrized so that {\em any} surface cuts could be obtained.

Lines were smoothened by fitting cuts to fourth-order polynomials
.

\bibliography{temperaturebib.bib}

\end{document}